\documentclass[10pt,english,onecolumn,floatfix,showpacs,amsmath,amssymb,aps]{revtex4}
\usepackage[T1]{fontenc}
\usepackage[latin1]{inputenc}
\usepackage{graphicx}
\makeatletter

\providecommand{\tabularnewline}{\\}

\usepackage{color}

\makeatletter

\usepackage{bm}

\usepackage{setspace}

\usepackage{color}

\makeatother

\usepackage{babel}
\makeatother
\begin{document}

\title{Elastic turbulence in von Karman swirling flow between two disks.}

\author{Teodor Burghelea}

\affiliation{Department of Mathematics, University of British Columbia, V6T1Z2
BC, Canada}

\author{Enrico Segre}

\affiliation{Department of Physical Services, Weizmann Institute of Science, Rehovot,
$76100$ Israel}

\author{Victor Steinberg}

\affiliation{Department of Physics of Complex Systems, Weizmann Institute of Science,
Rehovot, $76100$ Israel}

\date{\today }

\begin{abstract}
We discuss the role of elastic stress in the statistical properties
of elastic turbulence, realized by the flow of a polymer solution
between two disks. The dynamics of the elastic stress are analogous
to those of a small scale fast dynamo in magnetohydrodynamics, and
to those of the turbulent advection of a passive scalar in the Batchelor
regime. Both systems are theoretically studied in literature, and
this analogy is exploited to explain the statistical properties, the
flow structure, and the scaling observed experimentally. The following
features of elastic turbulence are confirmed experimentally and presented
in this paper:\\
 \textbf{(i)} the rms of the vorticity (and that of velocity gradients)
saturates in the bulk of the elastic turbulent flow, leading to the
saturation of the elastic stress.\\
 \textbf{(ii)} the rms of the velocity gradients (and thus the elastic
stress) grows linearly with $Wi$ in the boundary layer, near the
driving disk. The rms of the velocity gradients in the boundary layer
is one to two orders of magnitude larger than in the bulk.\\
 \textbf{(iii)} the PDFs of the injected power at either constant
angular speed or torque show skewness and exponential tails, which
both indicate intermittent statistical behavior. Also the PDFs of
the normalized accelerations, which can be related to the statistics
of velocity gradients via the Taylor hypothesis, exhibit well-pronounced
exponential tails.\\
 \textbf{(iv)} a new length scale, i.e the thickness of the boundary
layer, as measured from the profile of the rms of the velocity gradient,
is found to be relevant for the boundary layer of the elastic stresses.
The velocity boundary layer just reflects some of the features of
the boundary layer of the elastic stresses (rms of the velocity gradients).
This measured length scale is much smaller than the vessel size.\\
 \textbf{(v)} the scaling of the structure functions of the vorticity,
velocity gradients, and injected power is found to be the same as
that of a passive scalar advected by an elastic turbulent velocity
field.
\end{abstract}


\maketitle
\tableofcontents{}


\section{Introduction}

Long polymer molecules added to a fluid make it elastic and capable
to store stresses that depend on the history of deformation, thereby
giving the medium a memory~\cite{bird}. As was recently demonstrated,
elastic stress can overcome the dissipation caused by relaxating polymer
molecules, and can cause complicated and irregular motion in curvilinear
shear flows~\cite{Nat,NJP}. The discovery of elastic turbulence,
a random flow in dilute polymer solutions at arbitrary low Reynolds
numbers, $Re$, calls for quantitative studies of all aspects of the
phenomenon, both experimentally and theoretically.

Although the notion of turbulence itself is widely used in scientific
and technical literature, there is no commonly accepted definition.
Turbulent flow is usually identified by its main features. Turbulence
implies fluid motion in a broad range of temporal and spatial scales,
so that many degrees of freedom are excited in the system. There are
no intrinsic characteristic scales of time and space in the flow,
except for those restricting the excited temporal and spatial domains
from above and below. Turbulent flow is also usually accompanied by
a significant increase in momentum and mass transfer. That is, flow
resistance and rate of mixing in a turbulent flow become much higher
than those would be in an imaginary laminar flow with the same Reynolds
number.

Random flows of dilute polymer solutions too, cause a sharp growth
in flow resistance, exhibit an algebraic decay of velocity power spectra
over a wide range of scales, and provide a way for effective mixing~\cite{Nat,NJP,mix}.
These properties are analogous to those of hydrodynamic turbulence.
Such formal similarities give a reason for calling such flows \char`\"{}elastic
turbulence\char`\"{}. However, the similarities do not imply that
the physical mechanism that underlies the two kinds of random motion
is the same. Indeed, in contrast with inertial turbulence at high
$Re$, which occurs due to large Reynolds stresses~\cite{landau},
large elastic stress is the main source of non-linearity and the cause
of elastic turbulence in low $Re$ flows of polymer solutions~\cite{stretch}.
One can suggest that in a random flow driven by elasticity, the elastic
stress tensor $\bm{\tau}_{p}$ should be the object of primary importance
and interest, and that it may be appropriate to view elastic turbulence
as turbulence of the $\bm{\tau}_{p}$ field. It would then be more
relevant and instructive to explore the spatial structure and the
temporal distribution of this field, but, unfortunately, currently
no experimental technique allows a direct local measurement of it
in a turbulent flow. On the other hand, properties of the $\bm{\tau}_{p}$
field in a boundary layer can be inferred from measurements of injected
power, whereas its local properties can be evaluated from measurements
of spatial and temporal distributions of velocity gradients.

In this regard, we would like to point out the deep similarity between
elastic turbulence and the small-scale turbulent magneto-dynamo discussed
theoretically by~\cite{Volodya2}, that will be further clarified
in the next Section. Both the small-scale magneto-dynamo~\cite{schekochihin}
as well as elastic turbulence arise when the flow is three-dimensional,
random in time and spatially smooth. Analogously to elastic turbulence,
the small-scale magneto-dynamo exhibits a viscosity-dominated, low
$Re$ turbulence that is much faster than the large-scale one. The
small-scale turbulent dynamo is produced by the random stretching
of nearly frozen-in magnetic field lines by the ambient velocity field,
which is random and spatially smooth, with a growth rate of the order
of the inverse viscous eddy-turnover time. The stretching of the magnetic
field amplifies exponentially the field strength, at the rate of turbulent
stretching, and produces dynamically significant fields before the
large-scale fields can grow appreciably~\cite{schekochihin}. The
advection of the field then spreads the magnetic energy over a sub-viscous
range of scales and piles it up at the resistive scale. Below that
scale, diffusion dominates and causes an exponential decay of the
energy. This fast dynamo mechanism, due to stretching and folding,
generates magnetic field lines which remain straight and fairly well-aligned.
Similarly, passive scalar mixing in the Batchelor regime gives rise
to the so-called collinear anomaly~\cite{chertkov95} (see further
for details). The fundamental difference between the fast small-scale
magneto-dynamo and the elastic turbulence lies in what happens when
the small-scale magnetic field becomes sufficiently strong for the
Lorentz force to react back on the flow. The different dissipation
mechanism in the two systems leads to a different saturation mechanism.
In the latter system, saturation is reached when the dynamo stretching
that produces the feedback on the velocity field and the homogeneous
linear damping balance, whereas this balance is missing in the former
case.

The microscopic phenomenon responsible for the macroscopic elastic
turbulent flow of dilute polymer solutions is the so-called coil-stretch
transition, discussed originally by Lumley~\cite{lumley} and then
elaborated by de Gennes~\cite{deGennes}. The essence of the coil-stretch
transition is as follows: all the polymer molecules are normally,
in absence of a flow gradient, coiled into compact and nearly spherical
tangles with radii much smaller than the total polymer length. In
the presence of random flow gradients, the polymer molecules are deformed
into ellipsoid-like tangles. The mean deformation is moderate as long
as the stretching rate of the flow is smaller than the inverse polymer
relaxation time. When the stretching rate exceeds the inverse polymer
relaxation time, most of the molecules are strongly stretched to sizes
up to the order of the total polymer length, since stretching is limited
only by non-linear effects and back reaction of the flow. There is
an abrupt change in the polymer conformation occurring at large stretching
rates, called the coil-stretch transition. The transition has remarkable
macroscopic consequences. While in the coiled state the influence
of polymers on the flow is negligible and the polymer solutions can
be treated as Newtonian fluids (without memory), in the stretched
state polymers produce an essential feedback on the flow, and the
properties of the solution become strongly non-Newtonian. Let us stress
that the coil-stretch transition can occur even at small Reynolds
numbers or at low level of hydrodynamic nonlinearity. The elasticity
of the polymers itself (which manifests itself above the coil-stretch
transition) is a source of nonlinearity. This explains how a chaotic
state, which requires a high level of nonlinearity, can arise even
at low $Re$ in elastic turbulence.

Another important characteristic of the elastic turbulence regime
to be investigated is the scaling of the velocity field. The power
spectrum of velocity fluctuations measured in the experiments~\cite{Nat,NJP}
shows a power-law behavior $f^{-\delta}$ with $\delta>3$, where
$f$ is the frequency. Since the power spectrum decays strongly with
the frequency (and thus with the wave number), the fluctuations of
the velocity and of the velocity gradients are both determined by
the integral scale, i.e.\ the size of the vessel. This explains why
elastic turbulence is essentially a smooth random flow, dominated
by the strong nonlinear interaction of few large-scale modes. A theoretical
explanation of the power velocity spectrum was provided by Fouxon
and Lebedev~\cite{Volodya,Volodya2}. The spectrum $k^{-\delta}$
with $\delta>3$ implies that elastic turbulence flow can be assimilated
to a Batchelor regime flow, i.e.\ a temporally random and spatially
smooth flow~\cite{taylorhypoth}. This type of a random flow appears
in hydrodynamic turbulence below the dissipation scale.

In a spatially smooth flow, velocity differences are proportional
to distances between two fluid particles. As a result, particles separate
or approach exponentially in time. The statistics of stretching and
contraction of the fluid element can be described by the formalism
of dynamical systems, using the notion of Lyapunov exponents that
are the exponential rates of divergence. The number of the Lyapunov
exponents is equal to space dimensionality, and they are ordered in
such a way that the first Lyapunov exponent is the largest and describes
the separation rate of two fluid particles. Based on the Lagrangian
approach, the so-called collinear anomaly was predicted (first, within
the Kraichnan model and then for the general case): a spatially smooth
flow locally preserves straight lines~\cite{chertkov95}. This anomaly,
translated into the spatial distribution of the passive scalar, suggests
that parallel strips are the major structural elements in the Batchelor
regime. The collinear anomaly is also the source of strong intermittency
and of the absence of scale invariance observed at scales larger than
the forcing scale~\cite{chertkov95}. This simple yet fundamental
feature was observed experimentally for a passive scalar in elastic
turbulence~\cite{mix,Nat,Teo}.

Such a Batchelor flow is therefore an efficient mixer due to the exponential
divergence of Lagrangian trajectories. This property was directly
confirmed experimentally in curvilinear channel flows of polymer solutions,
where elastic turbulence was excited~\cite{mix,Teo}. The passive
scalar (dye) mixing had there an exponential character along the channel.
Additionally, the Lagrangian technique was used (a) to predict the
exponential tails of the passive scalar distribution function in the
steady regime, as was later confirmed experimentally and numerically;
(b) to account for a complex interplay between Lagrangian stretching
and diffusion, thus showing that the PDF of the scalar dissipation
field has a stretched-exponential tail~\cite{falkovich}; (c) to
explain strong intermittency in the scalar decay problem~\cite{son,fouxon},
where the decay rate is exponential in time and the exponent saturates
to a constant for high-order moments. While the experiment~\cite{mix}
confirmed the exponent saturation, it contradicted some other conclusions
of the theory developed for infinite domains~\cite{son,fouxon}.
This observation motivated theorists to develop a new (and much-needed)
theory of mixing in finite vessels~\cite{chertkov2}. This theory
explained that long-time properties of scalar decay are determined
by the stagnation regions of the flow, in particular boundary layers.
A nontrivial dependence of the decay rate on diffusivity was predicted,
as remarkably confirmed by the experiment on mixing in elastic turbulence~\cite{Teo}.
These experimental discoveries and subsequent theoretical developments
show the need to readjust the thinking in fundamental fluid mechanics
(that mixing of highly viscous fluids is difficult) but also open
up a whole new world of practical applications.

Elastic turbulence gives the possibility to dramatically increase
mixing rates (and, correspondingly, rates of chemical reactions) of
polymer melts in industrial reactors and in small-scale flows (via
capillaries, for instance). Indeed, to mix fluids effectively one
needs irregular motion, which brings into near contact distant portions
of the fluids. This method of mixing begins to find applications in
microfluidics and biophysics~\cite{patent}. It is likely that Nature
itself uses the trick of small-scale mixing by elastic turbulence.
Biological discoveries and applications along this direction are likely
to follow.

In this paper we present experimental results concerning statistics
of the injected power and statistics and structure of the velocity
field in an elastic turbulent flow, in its bulk as well as near a
wall. By using the analogy between passive scalar advection in the
Batchelor regime and advection of elastic stresses in elastic turbulence,
we try to elucidate the role of the latter, particularly in connection
with the intermittency and skewness of the injected power statistics,
with the statistics and saturation of rms of vorticity (velocity gradients)
in the bulk, and with various properties of the velocity profile in
the elastic turbulent flow regime, such as scaling and properties
of the boundary layer.

Our experimental program addresses the following questions:

\begin{description}
\item [{{(i)}}] Does elastic stress saturate in the bulk as predicted
theoretically, and what is the saturation value?
\item [{{(ii)}}] What is the mechanism for pumping energy from the driven
plate into the bulk?
\item [{{(iii)}}] What is the probability distribution of the elastic
stresses?
\item [{{(iv)}}] What is the velocity profile in the cell, and does an
analog of a boundary layer exist?
\item [{{(v)}}] If the boundary layer exists, which length scale characterizes
its thickness, and what is the nature of this scale?
\item [{{(vi)}}] Are there similarities in the statistical and scaling
behavior with the problem of turbulent advection of a passive scalar?
\end{description}
The paper is organized as follows. We start the next section presenting
a hydrodynamic description of dilute polymer solution flows and derive
non-dimensional parameters to characterize the flows that follow from
the equations. The variation of one of these control parameters, responsible
for the elastic properties of a fluid, can lead to elastic instability
in the swirling flow between two disks, distinguished by the presence
of curvilinear trajectories. The stretching of polymer molecules,
leading eventually to the coil-stretch transition, is further discussed
in section~\ref{sub:Polymer-extension}. The theoretical criterion
for the elastic instability in different flows, together with its
experimental verification, are discussed in section~\ref{sub:Elastic-instability}.
How this instability is responsible for the scenario dubbed elastic
turbulence is addressed in section~\ref{sub:ElasticTheory}. The
experimental measurement techniques used to characterize the flow
are presented in section~\ref{sub:-Experimental-set-up}, and to
complete the basics, the rheometric properties of polymer solutions
used are given in section~\ref{sub:RheometricProperties}, while
previous experimental observations of elastic instability in the von
Karman flow are summarized in section~\ref{sub:CriticalWi-inVonKarman}
and compared to our findings. In section~\ref{sec:Properties-of-elastic}
we give a complete account of the results of the measurements. Different
aspects of the flow are investigated with different techniques: we
concentrate on the statistics of global power fluctuations in section~\ref{sub:Power-fluctuations};
on the topology of the flow below and above transition and its relation
with the observed statistics in~\ref{sub:Flow-structure}; on profiles
of velocity and velocity gradients, and the emergence of boundary
layers in section~\ref{sub:Velocity-and-boundary}. All the new experimental
facts are summarized and discussed in in section~\ref{sub:Discussion},
where the role of elastic stress, a recent theory of elastic turbulence,
and comparative considerations about elastic versus hydrodynamic turbulence
are made. Finally, in section~\ref{sec:Conclusions} the results
are summarized and conclusions are drawn.

\section{Hydrodynamical theory of elastic turbulence\label{sec:Hydrodynamical-theory}}

\subsection{Hydrodynamical description of dilute polymer solution flows\label{sub:Hydrodynamical-description}}

Solutions of flexible high molecular weight polymers are viscoelastic
liquids, and differ from Newtonian fluids in many aspects~\cite{bird}.
The most striking elastic property of polymer solutions is, probably,
the dependence of mechanical stresses on the history of the flow.
The stresses do not immediately become zero when the fluid motion
stops, but rather decay with some characteristic relaxation time,
$\lambda$, which can be well above a second. When a polymer solution
is sufficiently diluted, its stress tensor $\bm{\tau}$ can be divided
into two parts, $\bm{\tau}=\bm{\tau}_{s}+\bm{\tau}_{p}$. The first
term, $\bm{\tau}_{s}$, is defined by the viscosity of the Newtonian
solvent $\eta_{s}$ and by the rate of strain in the flow, $\bm{\tau}_{s}=\eta_{s}[\mathbf{\nabla}\cdot\mathbf{V}+(\mathbf{\nabla}\cdot\mathbf{V})^{T}]$.
The elastic stress tensor, $\bm{\tau}_{p}$, is due to the polymer
molecules, which are stretched in the flow, and depends on the history
of the flow. The equation of motion for a dilute polymer solution
has the form~\cite{bird}

\begin{equation}
\frac{\partial\mathbf{V}}{\partial t}+(\mathbf{V\cdot\nabla)V}=-\frac{\mathbf{\nabla}p}{\rho}+\left(\frac{\eta_{s}}{\rho}\right)\mathbf{\nabla}^{2}\mathbf{V}+\frac{\nabla\cdot\bm{\tau}_{s}}{\rho}\,\,,\label{eq:motion}\end{equation}

where $p$ is the pressure and $\rho$ is the density of the fluid.
One can see that $\bm{\tau}_{p}$ enters the equation of motion linearly.
The equation has a non-linear term, $(\mathbf{V\cdot\nabla)V}$, which
is inertial in its nature. The Reynolds number $Re$ defines the ratio
of this non-linear term to the viscous dissipative term, $\nu\mathbf{\nabla}^{2}\mathbf{V}$.
Therefore the degree of non-linearity of the equation of motion can
still be defined, even for a polymer solution, by the Reynolds number,
$Re=VL\rho/\eta_{s}$. Turbulence in fluids at high $Re$ is a paradigm
for strongly nonlinear phenomena in spatially extended systems~\cite{landau,Tritt}.

The simplest model incorporating the elastic nature of the polymer
stress tensor, $\bm{\tau}_{p}$, is a Maxwell type constitutive equation~\cite{bird}
with a single relaxation time, $\lambda$,

\begin{equation}
\mathbf{\bm{\tau}}_{p}+\lambda{\frac{D\bm{\tau}_{p}}{Dt}}=\eta_{p}[\mathbf{\nabla\mathnormal{\mathnormal{\cdot}}{V}+\mathnormal{(}\nabla\mathnormal{\cdot}{V}\mathnormal{)}^{\mathnormal{T}}}]\,\,.\label{eq:taupMaxwell}\end{equation}
 Here $D\bm{\tau}_{p}/Dt$ is the material time derivative of the
polymer stress, and $\eta_{p}=\eta-\eta_{s}$ is the polymer contribution
to the total viscosity $\eta$. An appropriate expression for the
time derivative $D\bm{\tau}_{p}/Dt$ has to take into account that
the stress is carried by fluid elements, which move, rotate and deform
in the flow. The translational motion implies an advection term $(\mathbf{V}\cdot\mathbf{\nabla})\bm{\tau}_{p}$,
while the rotation and deformation of the fluid particles leads to
contributions like $(\mathbf{\nabla}\cdot\mathbf{V})\bm{\tau}_{p}$
or $\bm{\tau}_{p}(\mathbf{\nabla}\cdot\mathbf{V})$~\cite{bird}.
Therefore, some nonlinear terms, in which $\bm{\tau}_{p}$ is coupled
to $\mathbf{V}$, appear in the constitutive relation along with terms
linear in them. A simple model equation for $D\bm{\tau}_{p}/Dt$,
which is commonly used for description of dilute polymer solutions,
is the upper convected time derivative,

\begin{equation}
\frac{D\bm{\tau}_{p}}{Dt}=\frac{\partial\bm{\tau}_{p}}{\partial t}+(\mathbf{V}\cdot\mathbf{\nabla})\bm{\tau}_{p}-\bm{\tau}_{p}(\mathbf{V}\cdot\mathbf{\nabla})-(\mathbf{\nabla}\cdot\mathbf{V})^{T}\bm{\tau}_{p}\,\,.\label{eq:upperConvected}\end{equation}
 The equations (\ref{eq:taupMaxwell},\ref{eq:upperConvected}) together
with the expression for $\bm{\tau}_{s}$ constitute the Oldroyd-B
rheological model of polymer solutions~\cite{bird}. One can see
that the nonlinear terms in the constitutive equation (Eqs.~(\ref{eq:taupMaxwell},\ref{eq:upperConvected}))
are all of the order $\lambda(V/L)\tau_{p}$. The ratio of those non-linear
terms to the linear relaxation term, $\bm{\tau}_{p}$, that represents
dissipation, is given by a dimensionless expression $\lambda(V/L)$,
called the Weissenberg number, $Wi$, and expresses the ratio of the
relaxation time to the characteristic flow time. The relaxation term
$\bm{\tau}_{p}$ is somewhat analogous to the dissipation term in
the Navier-Stokes equation.

One can expect mechanical properties of the polymer solutions to become
notably nonlinear at sufficiently large Weissenberg numbers. Indeed,
quite a few effects originating from the nonlinear polymer stresses
have been known for a long time~\cite{bird}. In the simple shear
flow of a polymer solution, there is a difference between normal stresses
along the direction of the flow and along the direction of velocity
gradient. At low shear rates, the first normal stress difference,
$N_{1}$, is proportional to the square of the shear rate. For curvilinear
flow lines, this gives rise to a volume force acting on the liquid
in the direction of the curvature, called the {}``hoop stress''.
Therefore, if a rotating rod is inserted in an open vessel with a
highly elastic polymer solution, the liquid starts to climb up on
the rod, instead of being pushed outwards by the centrifugal force.
This phenomenon is known as {}``rod climbing'', or {}``Weissenberg
effect''~\cite{bird}. Furthermore, in a purely elongational flow
the resistance of a polymer solution depends on the rate of extension
in a strongly nonlinear way. There is a sharp growth in the elastic
stresses when the rate of extension exceeds $1/\lambda$, that is
at $Wi>1$. As a result, the apparent viscosity of a dilute polymer
solution in elongational flow can increase by up to three orders of
magnitude~\cite{sridhar}. Both the Weissenberg effect and the growth
of the extensional flow resistance were most clearly observed in viscous
polymer solutions and in flows with quite low $Re$, when nonlinear
inertial effects are insignificant~\cite{bird}.

If there are two potential sources of nonlinearity in the hydrodynamic
equations for a dilute polymer solution (Eqs.~(\ref{eq:motion},\ref{eq:upperConvected})),
inertial and elastic, a natural question arises: how is the hydrodynamical
stability of a flow affected by each nonlinear contribution? With
respect to the values of two non-dimensional parameters, $Re$ and
$Wi$, four cases can be considered in viscoelastic hydrodynamics.
If both $Re$ and $Wi$ are smaller than unity, the flow is steady
and laminar. Then in the limit of negligible $Wi$ and sufficiently
large $Re$, the flow loses its stability only because of the inertial
nonlinearity, which is much larger than the viscous dissipation, and
it evolves into the regime of inertial or hydrodynamic turbulence
of a Newtonian fluid. This remains true as long as the relaxation
time $\lambda$ is sufficiently small and the value of $Wi$ remains
negligible even at the large local velocity gradients appearing in
turbulent flow. The case of both large $Re$ and $Wi$ is commonly
realized in the hydrodynamic turbulence of fluids with small additions
of long polymer molecules. In such situation the most striking effect
observed is that of turbulent drag reduction. The last and less discussed
case is that of large $Wi$ and negligibly small $Re$, and is the
subject of this paper.

At first, it is not trivial to realize that some kind of turbulent
flow, driven solely by the nonlinear elastic stresses, may exist in
the absence of any significant inertial effects, at low $Re$. An
important step in this direction was made about a decade and a half
ago, when purely elastic instabilities were experimentally identified
in the rotational flow between two plates~\cite{magda} and in the
Couette-Taylor (CT) flow between two cylinders~\cite{LMS}.

\subsection{Polymer extension in random flows\label{sub:Polymer-extension}}

Recently Lumley's theory~\cite{lumley} was revised, and a quantitative
theory of the coil-stretch transition of a polymer molecule in a 3D
random flow was developed~\cite{lebedev,chertkov}. The dynamics
of a polymer molecule are indeed sensitive to the motion of the fluid
at the dissipation scale, where the velocity field is spatially smooth
and random in time. On this scale, polymer stretching is determined
solely by the velocity gradient tensor, $\nabla_{i}V_{j}$, which
varies randomly in time and space: $\partial_{t}R_{i}=R_{j}\nabla_{j}V_{i}-R_{i}/\lambda(R)+\zeta_{i}$.
Here $R_{i}$ are the end-to-end vector and $R$ is the end-to-end
distance for the stretched polymer molecule, respectively; $\lambda(R)$
is the polymer relaxation time, which is $R$-dependent, and $\zeta_{i}$
is the thermal noise. At $R\ll R_{max}$, where $R_{max}$ is the
maximum end-to-end stretched polymer length, the linear regime of
polymer relaxation is characterized by the polymer relaxation time
$\lambda_{rel}=\lambda(\dot{\gamma}=0)$. For molecules significantly
more elongated than in the coiled state one can use, e.g., the FENE
model with $\lambda(R)=\lambda(\dot{\gamma}=0)(1-R^{2}/R_{max}^{2})$~\cite{bird}.
In a $3D$ random flow, $\nabla_{i}V_{j}$ has always an eigenvalue
with a positive real part, so that there exists a direction with a
pure elongation flow~\cite{batch}. The direction and the rate of
the elongation flow change randomly, as the fluid element rotates
and moves along the Lagrangian trajectory. If $\nabla_{i}V_{j}$ remains
correlated within finite time intervals, the overall statistically
averaged stretching of the fluid element will increase exponentially
fast in time. The rate of the stretching is determined by the maximal
Lyapunov exponent, $\alpha$, of the turbulent flow, which is the
average logarithmic rate of separation of two initially close trajectories.
The value of $\alpha$ is usually of the order of the rms of the fluctuations
of the velocity gradient, $\left(\frac{\partial V_{i}}{\partial r_{j}}\right)^{rms}\equiv\overline{\left(\frac{\partial V_{i}}{\partial r_{j}}\right)^{2}}^{1/2}$.

The stretching of a polymer molecule follows the deformation of the
fluid element surrounding it. As a result, the statistics of polymer
elongations in a random smooth flow depends critically on the value
of $\lambda\cdot\alpha$ (or equivalently $\lambda\left(\frac{\partial V_{i}}{\partial r_{j}}\right)^{rms}$),
which plays the role of a local Weissenberg number for a random flow,
$Wi^{\prime}$. According to the theory~\cite{lumley,lebedev}, the
polymer molecules should become greatly stretched, if the condition
$\lambda\cdot\alpha>1$ is satisfied. Thus, the coil-stretch transition
is defined by the relation $\lambda_{cr}\cdot\alpha=1$, similarly
to that in an elongation flow with the strain rate $\alpha$~\cite{deGennes}.
A somewhat surprising conclusion of the theory is that a generic random
flow is, in average, an extensional flow in every point, with the
rate of extension $\dot{\varepsilon}=\alpha$ and unlimited Henky
strain. A dramatic extension of the flexible polymer molecules in
the turbulent flow environment, inferred here from bulk measurements
of the flow resistance, has been recently confirmed by the direct
visualization of individual polymer molecules in a random flow~\cite{corinne}.

An appropriate way of describing the coil-stretch transition employs
the probability distribution function (PDF) of polymer elongations.
Below the coil-stretch transition, the distribution is peaked near
the equilibrium tangle size, determined by thermal fluctuations. Above
the coil-stretch transition, the distribution of polymer sizes is
changed, and becomes peaked at an extension of the order of the total
polymer length. The position of the peak is determined by an interplay
between the stretching rate, the polymer nonlinearity, and the feedback
of the polymers on the flow. In a random flow, the distribution has
an extended (power law) tail related to stretching events that overcome
polymer relaxation, causing an essential extension of the polymer
molecule. The characteristics of the PDF tails and their relation
to the Lagrangian flow dynamics were theoretically examined in the
work by Balkovsky, Fouxon, and Lebedev~\cite{lebedev}. According
to their recent theory, the tail of PDF of molecular extensions is
described by the power law $P(R_{i})\sim R_{i}^{-\beta-1}$, where
$\beta\sim(\lambda^{-1}-\alpha)$ in the vicinity of the transition.
A positive $\beta$ corresponds to the majority of the polymer molecules
being non-stretched. On the contrary, at $\beta<0$ a significant
fraction of the molecules is strongly stretched, and their size is
limited by the feedback reaction of the polymers on the flow~\cite{lebedev}
and by the non-linearity of the molecular elasticity~\cite{chertkov}.
Thus, the condition $\beta=0$ can be interpreted too as the criterion
for the coil-stretch transition in turbulent flows~\cite{lebedev}.

To verify this theoretical picture, experimental studies of polymer
stretching in a 3D random flow between two rotating disks were conducted
in macro- and micro-setups~\cite{stretch,corinne}. It was experimentally
shown that elastic turbulence as a macroscopic phenomenon is tied
to the coil-stretch transition of polymer molecules that occurs on
a microscopic level~\cite{corinne}. In the microscopic experiment~\cite{corinne},
the size distribution of DNA molecules in the elastic turbulent flow
was directly measured by optical visualization methods. The validity
of the criterion for the coil-stretch transition was quantitatively
verified in the experiment by the analysis of both the velocity field
in the Lagrangian frame (via statistics of the Lyapunov exponents,
which defines the average $\alpha$), and of statistics of single
molecule stretching in the same flow~\cite{corinne}. A good agreement
between theory and experiment was found, both concerning the shape
of the polymer size distribution and the functional form of its tails.

\subsection{Elastic instability and its experimental investigation\label{sub:Elastic-instability}}

During the past decade, the purely elastic instabilities in viscoelastic
fluids have been the subject of many theoretical and experimental
studies, which are partially reviewed in Ref.~\cite{Larson,Shaq}.
Purely elastic instabilities of viscoelastic flows occur at $Wi$
of order unity and vanishingly small $Re$. As a result of the instability,
secondary vortex flows are developed~\cite{LMS}, and the flow resistance
increases~\cite{magda}. The analysis confirmed that the nonlinear
mechanical properties of the polymer solution can indeed lead to a
flow instability, for which a simple mechanism was proposed. After
the pioneering work by Larson, Muller and Shaqfeh~\cite{LMS,LMS2},
purely elastic instabilities were also found in other shear flows
with curvilinear streamlines. Those included the flow between a rotating
cone and a plate and the Taylor-Dean flow~\cite{Larson,Shaq}. The
original mechanism proposed in Ref.~\cite{LMS2} was verified experimentally
in Ref.~\cite{Ours2}. The original theoretical analysis of Ref.~\cite{LMS2}
was refined, and more elaborate experiments were carried out. A few
new mechanisms of flow instability driven by nonlinear elastic stresses
were suggested for those geometries.

The most thorough and detailed studies were conducted on the Couette-Taylor
(CT) flow between two coaxial cylinders. In spite of the fact that
instabilities in viscoelastic fluids were studied for decades, the
purely elastic instability in CT flow was first investigated both
experimentally and theoretically rather recently~\cite{LMS,LMS2}.
The mechanism of the elastic instability in the CT system suggested
in Ref.~\cite{LMS2} is based on the Oldroyd-B model. The primary
flow in the CT system with the inner cylinder rotating (Couette flow)
is a pure shear flow in the $r\theta$-plane that generates a normal
stress difference $N_{1}\equiv\tau_{\theta\theta}-\tau_{rr}=2\eta_{p}\lambda(\dot{\gamma}_{r\theta})^{2}$
and a radial force $N_{1}/r$ per unit volume. Here $r,\theta$ are
the cylindrical coordinates in the plane perpendicular to the cylinder
axis, $\tau_{\theta\theta}$ and $\tau_{rr}$ are the components of
the polymer stress tensor $\bm{\tau}_{p}$, and $\dot{\gamma}_{r\theta}$
is the only non-zero component of the rate-of-strain tensor, $\dot{\bm{\gamma}}$.
A secondary flow in the CT flow includes regions of elongational flow
in the $r$-direction with $\dot{\epsilon}\equiv\partial v_{r}/\partial r\neq0$.
The radial extensional flow stretches the polymer macromolecular coils
in the $r$-direction, though it is a small perturbation on top of
the azimuthal stretching produced by the primary shear flow. However,
when stretched in the radial direction, the macromolecular coils become
more susceptible to the basic shear flow. The coupling between the
radial and shear flow leads to further increase of shear stresses.
Thus, it results in further stretching of the polymer in the $\theta$-direction
that generates an additional normal stress difference. So, the elastic
instability mechanism is based on the coupling between the perturbative
radial elongation flow and the strong azimuthal shear flow, which
results in a radial force. The latter reinforces the radial flow.
As pointed out in Ref.~\cite{Ours2}, this transition can be only
a finite amplitude (first order) transition. The corresponding criterion
of the instability is

\begin{equation}
K\equiv\frac{\eta_{p}}{\eta_{s}}\frac{d}{R_{1}}Wi^{2}=O(1).\label{eq:K_CT}\end{equation}

Here $d$ and $R_{1}$ are the gap and the inner cylinder radius,
respectively. The Weissenberg number is defined here as $Wi=\lambda\Omega R/d$,
where $\Omega$ is angular velocity of the rotating inner cylinder.
(It was termed as the Deborah number in some of the original texts
\cite{LMS2,Ours2,EPL}). The elastic instability occurs when the parameter
$K$ exceeds a certain threshold value~\cite{LMS2}. This criterion
is valid at sufficiently small values of the polymer viscosity ratio,
$\eta_{p}/\eta_{s}$, and in the limit of small gap ratio $d/R_{1}$.
A more general expression is given in Ref.~\cite{LMS2}. When both
ratios are fixed, the elastic instability is defined only by the critical
Weissenberg number $Wi_{c}$.

As was suggested in Ref.~\cite{EPL}, where the CT flow was discussed,
there is some analogy between flow transitions driven by elasticity
and by inertia. The inertially driven Taylor instability occurs at
constant Taylor number~\cite{landau,Tritt}, $Ta={\frac{d}{R_{1}}}Re^{2}$,
while the elastic instability is controlled by the parameter $K$
of Eq.~(\ref{eq:K_CT})~\cite{LMS2,Ours2}. For the viscoelastic
case, the Weissenberg number appears to be analogous to the Reynolds
number. The geometric parameter determining curvature, $d/R_{1}$,
enters the expressions for both $Ta$ and $K$. Scales of time and
velocity for the purely elastic flow transition are given by $\lambda$
and $d/\lambda$, respectively. As it was shown in Ref.~\cite{EPL}
they are analogous to $t_{vd}$, the viscous diffusion time defined
as $t_{vd}=d^{2}/\nu$, and to $d/t_{vd}$, which provide characteristic
scales of time and velocity for the inertially driven flow transitions.

Beyond the analogy, there are important differences between the flow
transitions driven by inertia and by elasticity. It is an experimental
fact that \textit{any} laminar flow of a Newtonian fluid becomes unstable
at sufficiently high $Re$, and all high Reynolds number flows are
turbulent. That includes rectilinear shear flows, such as the Poiseuille
flow in a circular pipe, and the plane Couette flow, which are known
to be linearly stable at any $Re$. In contrast to it, purely elastic
flow instabilities in shear flows have been observed so far only in
systems with curvilinear stream lines. All these instabilities are
supposed to be driven by the hoop stress, which originates from the
normal stress differences. Recently, the authors of Ref.~\cite{vansaarlos}
presented theoretical arguments and numerical evidence for nonlinear
instability in viscoelastic plane Couette flow. This theory is waiting
its experimental verification and still remains quite controversial.

The difference between the inertial and elastic instabilities originates,
of course, from the distinct governing equations. There are, however,
some purely practical factors that can explain rather well the lack
of observations of purely elastic flow transitions in rectilinear
shear flows. Inertial instabilities in rectilinear shear flows of
Newtonian fluids occur at quite high Reynolds numbers. These are typically
about two orders of magnitude higher than $Re$ at which curvilinear
shear flows with large gap ratios become unstable. A priori, one may
suggest that rectilinear and curvilinear shear flows would have a
similar relation between $Wi$ at thresholds of the purely elastic
flow instabilities as well. The problem is that, while it is rather
easy to generate high $Re$ flows with Newtonian fluids of low viscosity,
it is usually impossible to reach the corresponding high values of
$Wi$ in shear flows of elastic polymer solutions. That is, there
are always rather severe practical limitations restricting nonlinearity
in elastic polymer stresses in shear flows, such as rapid polymer
degradation at high $Wi$.

A purely elastic instability, apparently due to a similar mechanism,
also occurs in other rotational shear flows, namely concentric cone-and-plate
and plate-and-plate geometries, which are common geometrical configurations
for rheological tests. In these cases the secondary flow is also driven
by a \char`\"{}hoop stress\char`\"{} and by the coupling between the
primary shear flow and secondary elongation flow in the axial direction.
Historically, a transition in the rheological measurement was first
observed experimentally in these geometries~\cite{magda}, and Magda
and Larson suggested that it had to be ascribed to an elastic flow
instability. Complete sets of numerical and experimental studies of
the elastic instability in these flows for small aspect ratio vessels
were carried on in Ref.~\cite{McK,oztekin,McKinley}. Due to the
more complicated flow structure than in CT flow, no analytic expression
for the stability criterion is available in either case. However,
some features common for all rotational shear flows can be pointed
out. For all systems with a small aspect ratio, the threshold of the
elastic instability depends on the aspect ratio that is defined differently
in each case. The instability criterion also depends on the viscosity
ratio, though the functional form of this dependence was found only
in the CT flow. In all systems, the most unstable mode is non-axisymmetric
and oscillatory and the transition is discontinuous or of the first
order (inverse bifurcation).

\subsection{Theory of the elastic turbulence and role of the elastic stress\label{sub:ElasticTheory}}

Let us consider now the limiting case of $Wi\gg1$ and $Re\ll1$.
Along with the apparent similarity in phenomenology of both elastic
and inertial turbulence, there are also many very important distinctions.
One similarity is reflected by the fact that the Weissenberg number,
which acts as the control parameter in elastic turbulence, depends
on the system size and on the fluid viscosity, like the Reynolds number
in hydrodynamic turbulence. Indeed, the scaling of the main characteristics
of the elastic turbulence with the system size, fluid velocity, and
viscosity is distinctly different from hydrodynamic turbulence. For
instance, it has been shown that with enough elasticity, one can excite
turbulent motion at arbitrary low velocities and in arbitrarily small
vessels~\cite{Nat,NJP}. An obvious reason for the difference in
scalings in the two scenarios is that the physical mechanisms which
underlie the two kinds of turbulent motion are themselves different.
As it is well known, the high flow resistance in inertial turbulence
is due to large Reynolds stresses. The Reynolds stress tensor is defined
as the average value of $\rho\left\langle V_{i}V_{j}\right\rangle $,
where $V_{i}$ and $V_{j}$ are different components of the flow velocity.
In the case of elastic turbulence, the Reynolds stresses are quite
small, since $Re$ is low. It follows that the high flow resistance
in elastic turbulence can only be due to the large elastic stresses,
$\bm{\tau}_{p}$~\cite{stretch}.

As we have already pointed out, the statistical properties of a turbulent
elastic flow and the increase in the contribution of the elastic stresses
to the flow resistance, are associated with the significant polymer
stretching in the random flow. To support this theoretical idea, some
experimental studies of polymer stretching in a 3D random flow between
two plates were conducted in macro- and micro-setups~\cite{stretch,corinne}.
The first experiment was performed in a swirling flow set-up with
the high aspect ratio $R/d=1$, where $R=30$ mm, using a very viscous
solvent with $\eta_{s}=1.36$ Pa$\cdot$s to suppress inertia~\cite{stretch}.
By experimental analysis and estimates of the contributions of the
Reynolds, viscous, and elastic stresses to the shear stress at the
upper plate, we found that the Reynolds stress accounts for less than
$0.1\%$ of the whole stresses, and that the viscous stress is rather
constant. So, as a result of a secondary random 3D flow superimposed
on an applied primary shear flow between two plates, the polymer contribution
to the shear stress increases as much as 170 times. If one assumes
a linear elasticity of the flow-stretched polymer molecules (polyacrylamide,
in this case), then the elastic stress causes an extension by 13 times~\cite{stretch}.

The molecular theory of polymer dynamics tells that $\bm{\tau}_{p}$
is proportional to the polymer concentration, $n$, and to the average
polymer conformation tensor, $\tau_{p,ij}\sim n\left\langle R_{i}R_{j}\right\rangle $~\cite{bird}.
Then, a significant increase in the contribution of $\bm{\tau}_{p}$
to the flow resistance in the elastic turbulence is associated, on
a microscopic level, with significant polymer stretching in a random
flow~\cite{lumley,lebedev,chertkov}.

The crucial theoretical step towards the description of elastic turbulence
relates the dynamics of $\bm{\tau}_{p}$ to the dynamics of a vector
field with a linear damping~\cite{Volodya,Volodya2,chertkov1}. The
elastic stress tensor can be written as uniaxial, i.e.\ $\tau_{ik}=B_{i}B_{k}$,
if the contribution of thermal fluctuations and polymer nonlinearity
to $\bm{\tau}_{p}$ can be neglected~\cite{Volodya}. Then, one can
write an equation for $B_{i}$ in a form which is similar to the equation
for magnetic field in magneto-hydrodynamics (MHD)~\cite{Volodya2}.
Thus, in the case of a viscoelastic flow, one gets: \begin{equation}
\partial_{t}\mathbf{B}+(\mathbf{V}\cdot\mathbf{\nabla})\mathbf{B}=(\mathbf{B\mathnormal{\cdot}\nabla)V}-\mathbf{B}/\lambda.\label{eq:visoelastic_withB}\end{equation}
 The difference with MHD lies only in the relaxation term that replaces
the diffusion term. As already mentioned in the Introduction, elastic
turbulence is analogous to a fast small-scale viscosity-dominated
magneto-dynamo, which realizes a sort of low Reynolds number Batchelor
regime of turbulence. The latter results from the random stretching
of the (nearly) frozen-in magnetic field lines by an advecting random
flow. Numerical simulations of the former show features very similar
to those observed in elastic turbulence~\cite{schekochihin}. Eq.~(\ref{eq:visoelastic_withB}),
complemented by the equation of motion, written for $Re\ll1$ as \begin{equation}
\mathbf{\nabla}P=\rho(\mathbf{B\mathnormal{\cdot}\nabla\mathnormal{)}B}+\eta\mathbf{\nabla^{\mathnormal{2}}V},\label{eq:motion_B}\end{equation}
and by appropriate boundary conditions, reveals the elastic instability
at $Wi=Wi_{c}>1$~\cite{LMS2,Ours2}, where $Wi=\dot{\gamma}\lambda$
and $\dot{\gamma}$ is the shear rate. At $Wi>Wi_{c}$ the instability
eventually results in chaotic, statistically stationary dynamics.
According to Ref.~\cite{Volodya}, a statistically steady state occurs
due to the back reaction of stretched polymers (or the elastic stresses
in Eq.(\ref{eq:motion_B})) on the velocity field, which leads to
a saturation of $\bm{\tau}_{p}$ even for linear relaxation. On the
other hand, the velocity gradients become smaller with decreasing
scale. This means that large-scale fluctuations (of the vessel size)
are dominant in the flow, and the dynamics are determined by a non-linear
interaction of modes on scales of the order of the system size. Assuming
that viscous and relaxation dissipative terms in Eqs.~(\ref{eq:visoelastic_withB})
and (\ref{eq:motion_B}) are of the same order, one gets that the
velocity gradients on the small scale, being fixed by the stationarity
conditions for statistics, are of the order of $\lambda^{-1}$~\cite{Volodya}.
So, the elastic stress can be estimated as

\begin{equation}
\tau_{p}=B^{2}\sim(\eta/\rho)\nabla_{i}V_{j}\sim\frac{\eta/\rho}{\lambda}.\end{equation}
 The fluctuating velocity vector and the stress tensor field (or $\mathbf{B}$)
can both be decomposed into large, $\mathbf{V}$ and $\mathbf{B}$
, and small, $\mathbf{v}$ and $\mathbf{B}^{\prime}$, scale components.
The analysis of the equations for the small-scale fluctuations of
both fields leads to a power-like decaying spectrum for the elastic
stress, which in spherical representation looks like $\langle B_{i}^{\prime}B_{j}^{\prime}\rangle\sim F(k)\sim k^{-\delta}$,
where $\delta>3$. It is clear that the field $B_{i}^{\prime}$ (and
so field $B_{i}^{\prime}B_{j}^{\prime}=\nabla_{i}v_{j}$) is the passive
field in the problem. The mechanism of stretching and folding of the
elastic stress field by a random advecting flow, with the key dynamo
effect and homogeneous attenuation of the stress field~\cite{chertkov1},
is rather general. It leads to the power-law spectrum for small-scale
fluctuations of $\mathbf{B}^{\prime}$ in a chaotic flow, and is directly
related to the classical Batchelor regime of mixing~\cite{batch},
where a passive scalar is advected by a flow which is spatially smooth
and random in time. A linear relation between the small-scale fluctuations
of the fields $\mathbf{B}^{\prime}$ and $\mathbf{v}$ allows one
to establish the power-law spectrum of for the velocity too, which,
in spherical representation looks again like $E(k)\sim k^{-\delta}$,
where $\delta>3$. This agrees well with the experimental values $\delta=3.3\div3.6$~\cite{Nat,NJP,thesis,my}.
Since the velocity spectrum decays faster than $k^{-3}$, the powers
of both the velocity and of the velocity gradients are determined
essentially by the vessel size; therefore elastic turbulent flow is
spatially smooth, strongly correlated on the global scale, and random
in time. That is also the main feature of the Batchelor regime, in
which smoothness in space and randomness in time flow are observed
below the dissipative scale~\cite{batch}.

We would like to emphasize that hydrodynamic turbulence instead exhibits
spatially rough flow on all scales above the viscous cut-off, and
cannot be approximated by a linear velocity field. Moreover, in contrast
to hydrodynamic inertial turbulence, the algebraic decay of the power
spectrum in elastic turbulence is not related to the energy cascade
nor to any conservation law, since the main energy dissipation occurs
at the largest scales. The mechanism leading to the algebraic power
spectra for the elastic stress and the velocity in this case is related
to the advection and linear decay accompanied by stretching of the
fluid element carrying the elastic stress. The fast decay of the fluctuation
power spectrum with $k$ implies a velocity field, where the main
contribution to deformation and stirring (stretching and folding)
at all scales comes from a randomly fluctuating flow field at the
largest scale of the system. The suggested mechanism for the generation
of small scale (high $k$) fluctuations in the elastic stress is the
advection of the fluid (which carries the stress) by this fluctuating
large scale velocity field. The theory considers the elastic stress
tensor being passively advected in a random velocity field, just as
the passively advected vector $\mathbf{B}^{\prime}$ in the magnetic
dynamo theory~\cite{Volodya2}.

The smoothness of the velocity field in elastic turbulence was experimentally
tested by investigating the shape of the cross-correlation functions
of the velocity field~\cite{taylorhypoth}. It was found there that
the second order spatial derivative of the velocity field is roughly
an order of magnitude smaller than the first one. Thus, a linear velocity
field (locally uniform rate of strain) is a good local approximation
for the velocity field in elastic turbulence~\cite{taylorhypoth}.

\section{Experimental\label{sec:Experimental}}

\subsection{Experimental set-up and procedure.\label{sub:-Experimental-set-up}}

\begin{figure}
\includegraphics[width=13cm]{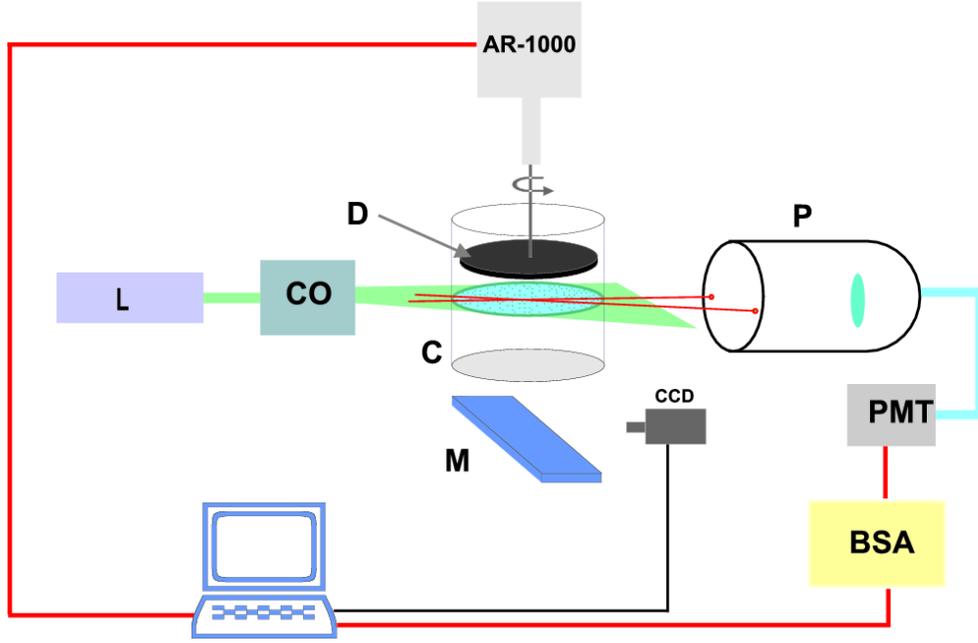}

\caption{Overview of the swirling flow apparatus: \textbf{C}-fluid container,
\textbf{D}-rotating disk, \textbf{AR-1000}-rheometer, \textbf{CO}-cylindrical
optics, \textbf{L}-laser, \textbf{M}-plane mirror, \textbf{CCD}-digital
camera, \textbf{P}-LDV probe, \textbf{BSA}-burst spectrum analyzer,
\textbf{PMT}-photomultiplier.}

\label{f.1}
\end{figure}

The experiments were conducted in the apparatus schematically shown
in Fig.\ref{f.1}. It consists of a stationary cylindrical container
\textbf{C,} with radius $R_{c}$, mounted on the basis of a commercial
rheometer (AR-1000 from TA Instruments) and a rotating disk \textbf{D,}
with radius $R_{d}$, mounted concentrically on the shaft of the rheometer.
In order to check the sensitivity of the results with respect to the
geometrical aspect ratio and to explore a broad range of Weissenberg
numbers, two versions of this setup were used: the first had $R_{c}=2.2cm$,
$R_{d}=2cm$ and height $d=1cm$, and the second had $R_{c}=4.9cm$,
$R_{d}=4.7cm$ and $d=1cm$. The two setups will be further referred
to as setup $1$ and setup $2$. The cylinder \textbf{C} was filled
with fluid just up to the disk, whose lower side only was wet. In
order to provide thermal stabilization, the whole rheometer was placed
in a thermally isolated box with through flow of temperature controlled
air. The rheometer mechanism allowed precise control (within 0.5\%)
of the angular velocity $\Omega$ of the disk, and simultaneous measurements
of the applied torque $T$ ($\Omega$-forcing mode), or conversely
-- control (within 0.5\%) of $T$ and measurements of $\Omega$ ($T$-forcing
mode). The average shear stress at the upper plate, $\tau_{w}$, was
calculated using the equation $T=\tau_{w}\int rdS$, that gave $\tau_{w}\equiv3T/(2\pi R_{d}^{3})$.
(The integration is carried over the upper plate surface).

The moment of inertia of the shaft of the rheometer was $I_{s}\approx14\mu N\cdot m\cdot s^{2}$
and that of the upper plate, $I_{d}$, was about $61\mu N\cdot m\cdot s^{2}$
for setup $1$, and about $84\mu N\cdot m\cdot s^{2}$ for setup $2$.
The accuracy of the angular speed measurements in constant torque
mode is about $2\%$ and the accuracy of the torque measurements in
the constant speed mode is about $1\%$. One has to point out here
that the small rate of the fluctuations of the angular velocity is
not a sufficient criterion to have a constant speed forcing. In the
$\Omega$-mode, for instance, $(I_{s}+I_{d})\frac{\partial\Omega}{\partial t}$
should also be much smaller than the typical values of torque, $T$.

To obtain a detailed characterization of both the spatial structure
and the temporal evolution of the flow-fields, three different experimental
techniques have been alternatively used: particle image velocimetry
(PIV), particle tracking velocimetry (PTV) and laser Doppler velocimetry
(LDV). The cup \textbf{C} was machined from a single piece of optically
clear perspex, to allow side and bottom view. The cup had a circular
inner and a square outer horizontal cross-section (i.e.\ flat sides)
in order to insure a distortion free illumination, and was attached
to the rheometer base in such a way that view from below was possible
through a mirror $\mathbf{M}$, tilted by $45^{\circ}$. As flow tracers,
we have used $10\mu m$ fluorescent particles for the PIV and PTV
measurements, and $1\mu m$ latex beads for the LDV measurements.
For PIV and PTV the system was illuminated laterally by a thin ($30\mu$m
in the center of the set-up and about $130\mu$m at the edges of the
set-up) laser sheet through the transparent walls of the fluid container
at the middle distance between the plates. The laser sheet was generated
by passing a laser beam delivered by a $300mW$ argon-ion laser, \textbf{L},
through a block of two crossed cylindrical lenses, \textbf{CO}, mounted
in a telescopic arrangement. The mirror \textbf{M} was used to relay
images of the flow to a CCD camera (PixelFly digital camera from PCO).
The camera was equipped with a regular video lens and mounted horizontally
near the rheometer (Fig.\ref{f.1}). Images were acquired with 12
bit quantization and resolution of $640\times512$ pixels at up to
$25$ frame/sec and $1280\times1024$ at up to $12.5$ frame/sec.

The main tool for the investigation of the flow was the digital PIV
technique. We acquired series of up to $2000$ pairs of flow images
with the camera. A custom development of the camera control software
allowed us to adjust the time delay between consecutive PIV images
in relation to the local flow velocity (in order to keep the mean
particle displacement in the range $5-15$ pixels). Corresponding
to low values of the angular velocity of the upper disk (up to about
$0.4rad/s$) the time delay was $540ms$ and then, gradually decreased
down to $40ms$. The total data acquisition time was always longer
than the Eulerian correlation times of the velocity (about $24$ times
in the fully developed random regime). Time series of velocity fields
were obtained by a multi-pass PIV algorithm~\cite{taylorhypoth}.
The accuracy of the method was carefully checked by running test experiments
with the solvent, in the same range of mean particle displacements
and at similar illumination conditions. Although the instrumental
error increases more or less linearly with the mean particle displacement,
it never exceeded $5\%$ of the mean displacement. The spatial resolution
of the PIV velocity measurements was $0.57mm$ in setup~$1$ and
$0.89mm$ in setup~$2$. By post-processing the velocity fields,
we obtained the profiles of the velocity components, fields of fluctuations
of each velocity component, spatial spectra of the velocity fluctuations,
velocity gradients and their fluctuations, structure functions of
gradients, and Eulerian velocity correlation functions. The space-time
measurements together with simultaneous global measurements of the
flow resistance provided a rather complete description of the different
flow regimes as a function of $Wi$~\cite{thesis,my,teoprl}.

We also examined Lagrangian properties of the flow~\cite{teo}. In
order to check the consistency of the PIV approach (particularly when
Lagrangian trajectories were measured), flow fields have been alternatively
investigated with the PTV technique. The first step of the PTV approach
is the accurate particle identification in each flow image, by correlation
with prototypical particle shapes, which can be either extracted from
flow images, or defined manually. Then, trajectories are reconstructed
by joining successive positions of the same particle on subsequent
images. This has been done by a search algorithm based on an initial
\char`\"{}guess\char`\"{} of the mean flow line.

Although the PIV and PTV tools are very suitable for the investigation
of the spatial properties of the flow, they provide a limited time
resolution and statistics. In order to overcome this limitation, the
LDV technique has alternatively been used. With the LDV technique,
time series of the fluid velocity are measured in a relatively small
volume (typically $100\mu m\times50\mu m\times50\mu m$). The setup
allowed measurements of one component of the velocity by LDV with
two crossing and frequency shifted beams. By appropriate positioning
and orientation of the beam crossing region, an azimuthal (longitudinal)
velocity component, $V_{\theta}$, could be measured at different
$r$ and $z$. Here $(r,\theta,z)$ are the standard cylindrical coordinates.
A commercial LDV system from Dantec Dynamics Inc.\ was used. Two
laser beams delivered by a probe (\textbf{P} in Fig.~\ref{f.1})
are aligned to cross each other at an angle of $\theta=10^{\circ}$
at some point in the flow. The light scattered in the backward direction
by the flow tracers present in the flow is collected by the probe
\textbf{P} and delivered to a photomultiplier, \textbf{PMT}, via a
fiber optic light guide. The PMT signal is processed by a real time
burst spectrum analyzer, \textbf{BSA}. The local flow velocity is
proportional to the Doppler shift of the signal. The error validated
velocity data are stored on a computer via a GPIB line.

Our experiments need to be classified according to meaningful values
of the non-dimensional control parameters $Re$ and $Wi$, which depend
among the rest on a global shear rate. In a swirling flow between
two plates, the shear rate is quite inhomogeneous over the fluid bulk,
even when the flow is laminar. Therefore, the choice of a representative
shear rate becomes somewhat arbitrary. We decided to consider the
simple expression $\Omega R/d$ as a characteristic shear rate, and
to define the Weissenberg number as $Wi=\lambda\Omega R/d$. The Reynolds
number was defined as $Re=\Omega Rd\rho/\eta$.

\subsection{Rheometric properties of polymer solutions\label{sub:RheometricProperties}}

The polymer used was polyacrylamide, PAAm (from Polysciences Inc.),
with a molecular weight $M_{w}=1.8\times10^{7}Da$, as used in Ref.~\cite{Nat,NJP,mix}.
Our preparation recipe was as follows: first we dissolved $0.9$ g
of PAAm powder and $3$ g of NaCl into $275$ ml of deionized water
by gentle shaking. NaCl was added to fix the ionic contents. Next
the solution was mixed for $3$ hours in a commercial mixer with a
propeller at a moderate speed. The rationale for this step is to cause
a controlled mechanical degradation of the longest PAAm molecules,
in order to \char`\"{}cut\char`\"{} the tail of the molecular weight
distribution of the broadly dispersed PAAm sample. In a solution with
a broad distribution of polymer molecular weights the heaviest molecules,
which are most vulnerable to mechanical degradation, bring the major
contribution to the solution elasticity, but may break in the course
of the experiment. This can lead to inconsistency of the experimental
results. We found empirically that the procedure of pre-degradation
in the mixer leads to substantial reduction of degradation during
the experiments and to substantial improvement of their consistency~\cite{NJP}.
Finally, $9$ g of isopropanol was added to the solution (to preserve
it from aging) and water was added up to $300$ g. The final concentrations
of PAAm, NaCl and isopropanol in the stock solution were $3000$ ppm,
$1\%$ and $3\%$, respectively. This master solution was used to
prepare $80ppm$ PAAm solution in a Newtonian solvent. The Newtonian
solvent was about $65\%$ saccharose in water~\cite{NJP}.

The rheological properties of the solvent and the polymer solution
were measured with two different rheometers: the AR-1000 from TA Instruments
around which the complete experimental setup was built, and a Vilastic
3 from Vilastic Scientific. The viscosity of the solvent was found
to be $\eta_{s}=114mPa\cdot s$ at $22^{o}C$, and that of the solution
$\eta=138mPa\cdot s$ at a shear rate of $2s^{-1}$. The polymer relaxation
time $\lambda$ was measured in oscillatory tests at different shear
rates, with $\dot{\gamma}$ ranging from $0.4s^{-1}$ to $3.6s^{-1}$.
In the limit of $\dot{\gamma}\rightarrow0$ the relaxation time was
$\lambda(0)=4.7$ s. The solution showed a clear \char`\"{}shear thinning\char`\"{}
as a function of the shear rate with the scaling $\lambda\sim\dot{\gamma}^{-\delta}$,
where $\delta\simeq0.3$, similar to what was found earlier for polymer
solutions with lower molecular weight PAAm samples~\cite{Ours2}.

\subsection{Experimental observation of elastic instability in von Karman swirling
flow\label{sub:CriticalWi-inVonKarman}}

The functional dependence of the elastic instability threshold on
the aspect ratio, $\epsilon\equiv d/R_{c}$, in a flow between two
plates was first obtained in the experiment by Magda and Larson~\cite{magda}.
They found an inverse dependence of the critical shear rate $\dot{\gamma}_{c}$,
or of the critical Weissenberg number, $Wi_{c}=\dot{\gamma}_{c}\lambda$,
on the aspect ratio, i.e.\  $Wi_{c}\propto\epsilon^{-1}$. The experiment
was conducted in the range of low values of $\epsilon$ between $0.024$
and $0.16$. In the later experiments similar results were obtained
in the same range of $\epsilon$ by McKinley and collaborators~\cite{McK,oztekin,McKinley}.
These results were found in reasonable agreement with numerical calculations.
However, these numerical results presented in $Wi_{c}-\epsilon$ coordinates
can be fitted by $Wi_{c}\propto\epsilon^{-0.470\pm0.006}$ in a wide
range of $\epsilon$ between $10^{-3}$ and $1$, in odds with the
the above mentioned dependence $Wi_{c}\propto\epsilon^{-1}$.

\begin{figure}[h]

\begin{centering}\includegraphics[width=12cm]{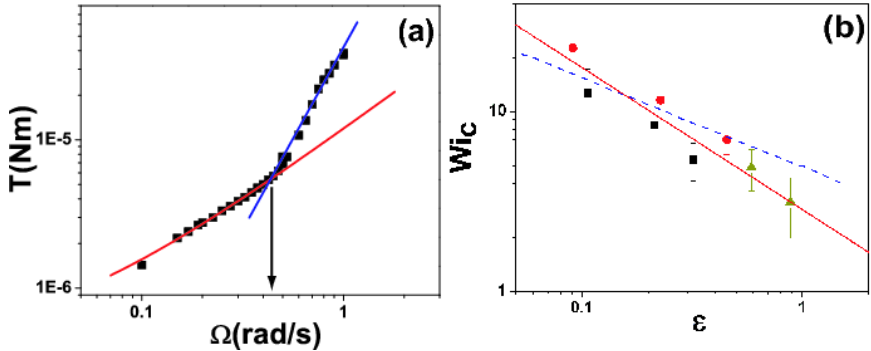} \end{centering}

\caption{Dependence of the onset of the primary elastic
instability on the geometric aspect ratio: (a) Torque as a
function of angular velocity for the aspect ratio $\epsilon=0.88$.
The full lines are linear fits. The arrow indicates the onset of
the primary elastic instability. (b) Critical Weissenberg number
for the onset of the elastic instability as a function of the
aspect ratio. The symbols are (color online): (black)
squares--$R_{c}=4.7cm$, (red) circles--$R_{c}=2.2cm$, (yellow)
triangles--$R_{c}=1.7cm$. The full line is a linear fit to the
data, $Wi_{c}=0.51\epsilon^{-0.79}$. The dashed line is the fit to
the numerical result from~\cite{oztekin},
$Wi_{c}=0.62+5.42{\epsilon}^{-0.47}$.} \label{f.2}
\end{figure}

We have investigated experimentally the dependence of the onset of
the elastic instability on the aspect ratio of our setup. By modifying
both the radius of the container, $R_{c}$, and the distance between
plates, $d$, the aspect ratio $\epsilon$ was varied by about 10
times for a rather wide range of $\epsilon$, between $0.091$ and
$0.882$. The onset values of the angular velocity, $\Omega_{c}$,
were obtained from the dependence of the averaged torque on the angular
velocity of the disk (Fig.~\ref{f.2}). By linear approximation of
this dependence on both sides of the transition point, the latter
can be determined within reasonable error bars (see Fig.~\ref{f.2}a).
The Reynolds number corresponding to the onset of the elastic instability,
$Re_{c}$, was of the order of unity in both setups, suggesting that
the transition is of purely elastic nature. The results of these measurements
together with other parameters are presented in Table~\ref{onsetvalues}.

\begin{table}[h]
\begin{centering}\centering \begin{tabular}{||c|c|c|c|c|c|c|c|c||}
\hline
${\bf {R_{c}(cm)}}$&
${\bf {d(cm)}}$&
${\bf {\epsilon}}$&
${\bf {\Omega_{c}(rad/s)}}$&
${\bf {\dot{\gamma}_{c}(s^{-1})}}$&
${\bf {\lambda(\dot{\gamma})_{c}(s)}}$&
${\bf {Wi_{c}}}$&
${\bf {Re_{c}}}$&
${\bf {\delta Wi_{c}}}$ \tabularnewline
\hline
1.7&
1&
0.59 &
0.64 &
1.09 &
4.485 &
4.9&
1.03 &
1.28 \tabularnewline
\hline
1.7 &
1.5 &
0.88 &
0.51&
0.58 &
5.42 &
3.13 &
1.23&
1.16 \tabularnewline
\hline
2.2 &
1 &
0.45&
0.82 &
1.8 &
3.85 &
6.95 &
1.72 &
1.19 \tabularnewline
\hline
2.2 &
0.5 &
0.23 &
0.85 &
3.74 &
3.1 &
11.6 &
0.89 &
2.45 \tabularnewline
\hline
2.2 &
0.2 &
0.091 &
0.88 &
9.68 &
2.33 &
22.54 &
0.37 &
4.94 \tabularnewline
\hline
4.7 &
0.5 &
0.106 &
0.45 &
4.23 &
2.98 &
12.62 &
1.0 &
4.6 \tabularnewline
\hline
4.7 &
1 &
0.213 &
0.5 &
2.35 &
3.56 &
8.37 &
2.24 &
1.33 \tabularnewline
\hline
4.7 &
1.5 &
0.32 &
0.4 &
1.25 &
4.3 &
5.39 &
2.7 &
1.29 \tabularnewline
\hline
\end{tabular}
\end{centering}

\caption{Onset values of the relevant physical quantities corresponding to
the primary elastic instability }

\label{onsetvalues}
\end{table}

We remark that the values of the critical Weissenberg number as a
function of the aspect ratio obtained from our experimental results
agree fairly well with the linear stability analysis presented in~\cite{oztekin,McKinley,McK},
as shown in Fig.\ref{f.2}b. On the other hand, the fit to the theoretical
functional dependence in a wide range of $\epsilon$ between $10^{-3}$
and 1, namely $Wi_{c}\propto\epsilon^{-0.47}$, is quite different
from that obtained from the experimental data: $Wi_{c}\propto\epsilon^{-0.79\pm0.16}$.
This discrepancy can be attributed to the larger aspect ratios experimentally
studied by us.

\section{Properties of elastic turbulence\label{sec:Properties-of-elastic}}

\subsection{Power fluctuations and statistics of the injected power\label{sub:Power-fluctuations}}

\begin{figure}
\includegraphics[width=12cm]{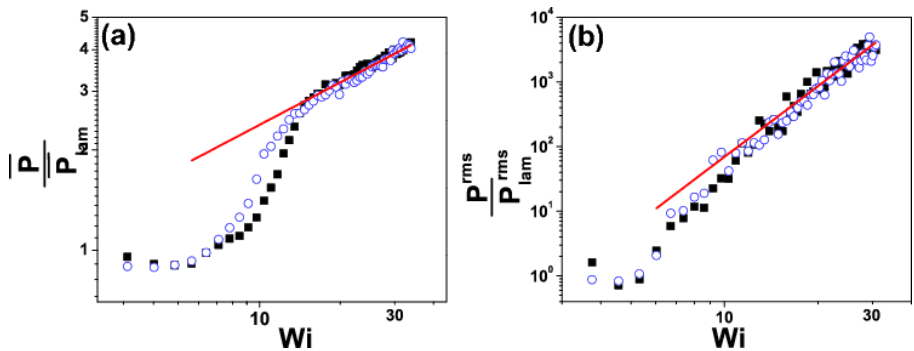}

\caption{(a) Dependence of the scaled average power, $\overline{P}/\overline{P}_{lam}$,
on the control parameter $Wi$: squares-increasing $Wi$, circles-decreasing
$Wi$. The full line is a guide for the eye, with slope $Wi^{0.49}$.
(b) Dependence of the scaled rms of the power fluctuations, $P^{rms}/P_{lam}^{rms}$
on $Wi$: squares-increasing $Wi$, circles-decreasing $Wi$. The
full line is a power law, $Wi^{3.2}$. Data were collected in set-up
$2$ in $\Omega$ forcing mode.}

\label{f.3}
\end{figure}

One of the main features of the transition to elastic turbulence is
the substantial growth of the flow resistance above the onset of the
instability~\cite{Nat,NJP}. In the case of a von Karman swirling
flow, a measure of the flow resistance is either the power $P_{\Omega}$
needed to spin the upper disk at constant angular speed $\Omega$,
or the power $P_{T}$ needed to spin the upper disk with constant
torque $T$ applied to the shaft of the rheometer. The injected power
is defined in any case as $P=T\cdot\Omega$. In our experiments, due
to the smallness of $Re<16$, the inertial contribution was always
low. The dependencies of the reduced average injected power, $\bar{P}/P_{lam}$
and of the rms of its fluctuations, $P^{rms}/P_{lam}^{rms}$, in a
wide range of the control parameter $Wi$ in the $\Omega$ mode are
presented in Fig.~\ref{f.3}. Here $P_{lam}$ and $P_{lam}^{rms}$
are the injected power and its fluctuations before the onset of the
elastic instability.

The data presented in Fig.~\ref{f.3} reveal three distinct flow
regimes. For low values of the control parameter $Wi$, the reduced
injected power is equal to unity. At $Wi\simeq Wi_{c}$, a primary
elastic instability occurs, resulting in a sharp increase of the injected
power and of the rms of its fluctuations. A further increase of the
control parameter causes the flow to evolve towards a scaling regime,
where both the reduced average and rms of fluctuations of the injected
power scale algebraically with $Wi$ (Fig.~\ref{f.3}): $\bar{P}/P_{lam}\propto Wi^{0.49\pm0.05}$
and $P^{rms}/P_{lam}^{rms}\propto Wi^{3.2\pm0.3}$.

\begin{figure}
\includegraphics[width=12cm]{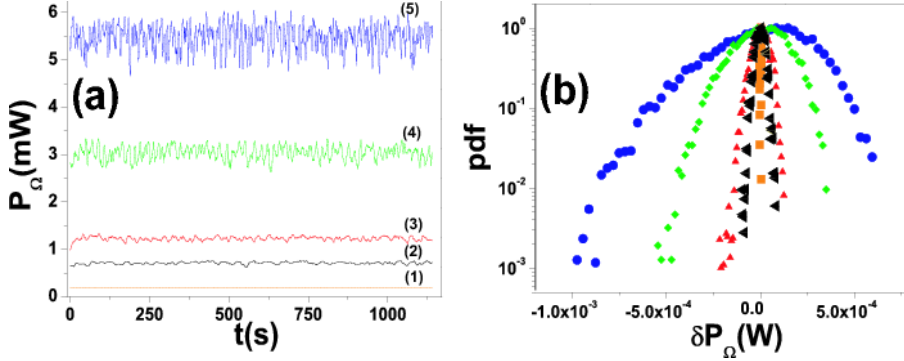}

\caption{(a) Time series (partially shown) of the injected power in constant
$\Omega$ forcing regime, for different $Wi$. The labels are: \textbf{(1)}-$Wi=5$,
\textbf{(2)}-$Wi=19$, \textbf{(3)}-$Wi=24$, \textbf{(4)}-$Wi=31.5$,
\textbf{(5)}-$Wi=40$. (b) PDF's of the power fluctuations in$\Omega$-forcing
regime for different $Wi$. The symbols are (color online): squares
(orange)-$Wi=5$, left triangles (black)-$Wi=19$, up triangles (red)
-$Wi=24$, diamonds (green)-$Wi=31.5$, circles (blue)-$Wi=40$. Data
were collected in setup 2.}

\label{f.4}
\end{figure}

\begin{figure}
\includegraphics[width=12cm]{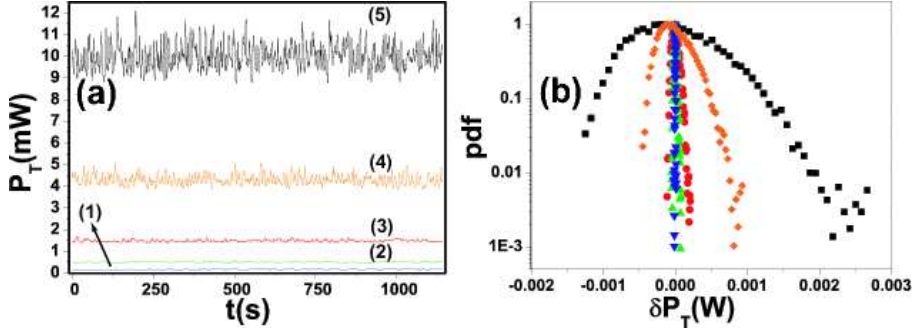}

\caption{(a) Time series (partially shown) of the injected power in constant
$T$ forcing regime for different $Wi$. The labels are: \textbf{(1)}-$Wi=5$,
\textbf{(2)}-$Wi=17$, \textbf{(3)}-$Wi=25$, \textbf{(4)}-$Wi=34$,
\textbf{(5)}-$Wi=46$. (b) PDF's of the power fluctuations in $T$-forcing
regime for different $Wi$. The symbols (color online) are: down triangles
(blue)-$Wi=5$, up triangles (green)-$Wi=17$, circles (red)-$Wi=25$,
diamonds (orange)-$Wi=34$, squares (black)-$Wi=46$. Data were collected
in setup 2.}

\label{f.5}
\end{figure}

\begin{figure}[h]

\begin{centering}\includegraphics[width=12cm]{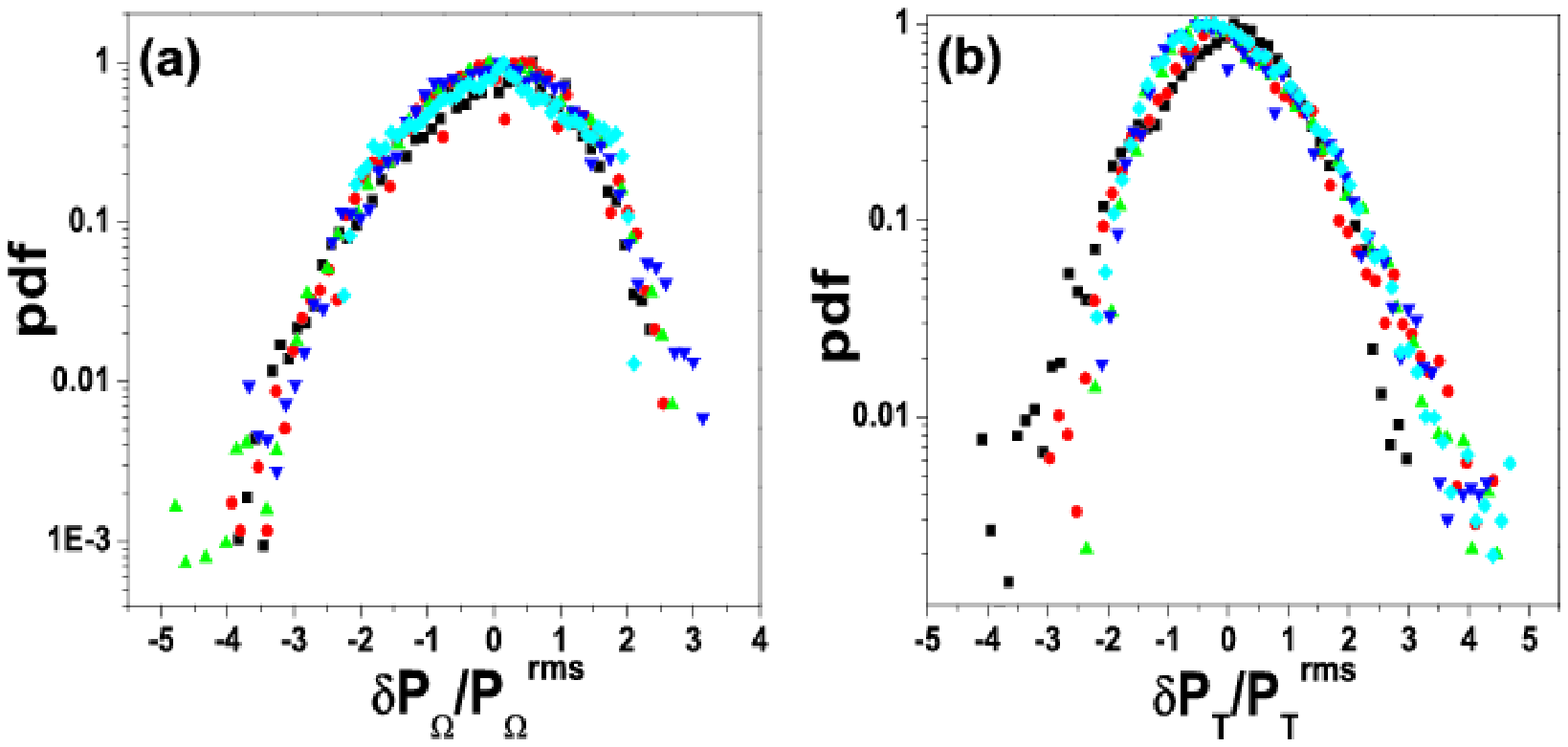} \end{centering}

\caption{(a) scaled PDFs of the injected power in constant $\Omega$ forcing
regime for different $Wi$. (b) scaled PDFs of the power fluctuations
in constant $T$ forcing regime for different $Wi$. The symbols (color
online) are: down triangles (blue)-$Wi=5$, up triangles (green)-$Wi=17$,
circles (red)-$Wi=25$, diamonds (light blue)-$Wi=34$, squares (black)-$Wi=46$.}

\label{f.6}
\end{figure}

The fluctuations of the injected power were measured for different
$Wi$ in the elastic turbulence regime in two modes: $\Omega$-forcing
is presented in Fig.~\ref{f.4} and $T$-forcing is presented in
Fig.~\ref{f.5}. For each value of $Wi$ the statistics of the power
fluctuations was collected on $180000$ data points evenly sampled
in time ($\Delta t\approx38ms$). The time series of the injected
power at different $Wi$ for both modes are presented in both figures.
In order to avoid mechanical degradation of the the polymer solution
during the long data acquisition times, separate experiments were
conducted for each value of the control parameter. The local (in time)
average values of the injected power did not change significantly
during the total data acquisition times, suggesting that no major
degradation occurred. For low values of the control parameter in the
laminar regime, ($Wi<Wi_{c}$), the power fluctuations are only due
to the instrumental noise. In the elastic turbulence regime, the PDFs
of the power fluctuations strongly deviate from Gaussian distributions
for both forcing modes. The PDFs in the $\Omega$-forcing mode have
a left side skewness, while in the $T$-forcing mode -a right side
skewness. Since the fluctuations of the injected power reflect fluctuations
of the elastic stress averaged over the upper plate, one can conclude
that the statistics of the elastic stress too must strongly deviate
from Gaussian, which is a signature of intermittency. The PDFs of
the reduced power fluctuations collapse on a single curve when normalized
by their maximum value (see Fig.~\ref{f.6}). Both the dependency
on $Wi$ of $\bar{P}/P_{rms}$ (Fig.~\ref{f.3}), and the skewness
of the PDF in elastic turbulence suggest an analogy with similar behavior
in hydrodynamic turbulence, though the reasons for the effects are
entirely different~\cite{pinton,cadot}.

\begin{figure}[h]

\begin{centering}\includegraphics[width=12.5cm]{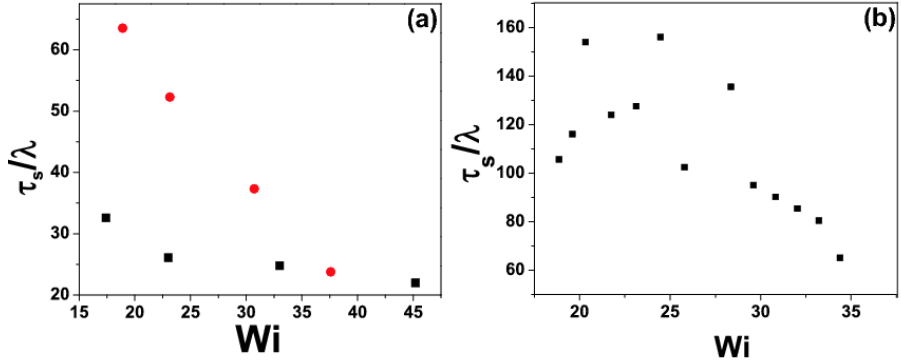} \end{centering}

\caption{(a) Normalized spike periodicity, $\tau_{s}/\lambda$, in time series
of the injected power vs.\ $Wi$: circles-constant $\Omega$ mode,
squares-constant $T$ mode. (b) Normalized spike periodicity, $\tau_{s}/\lambda$,
in time series of the azimuthal velocity vs.\ $Wi$. Data were collected
at $r=2R_{c}/3$, $z=d/2$ in setup $2$.\label{f.7} }
\end{figure}

We further examined the statistics of rare and short time scale events
(spikes) in the injected power series; namely we tried to quantify
the characteristic width of the spikes and the characteristic time
between them. To carry out this analysis, the following procedure
was applied. First, all local minima and maxima were detected in each
time series. Then only those minima (in the case of constant $\Omega$
forcing) and those maxima (in the case of constant $T$ forcing) that
contribute to the skewness of the corresponding PDFs, are selected,
i.e.\ those peaks which are larger than $2P^{rms}$. The values of
the average time $\tau_{s}$ between two spikes, normalized by the
relaxation time, are presented in Fig.~\ref{f.7}(a), as function
of $Wi$ in both $\Omega$ and $T$ modes.

We performed a similar analysis on the time series of the local azimuthal
velocity measured by the LDV technique, such as those shown in the
following in Fig.~\ref{velstat}. The values of the the average time
between two spikes, normalized by the relaxation time, are presented
in Fig.~\ref{f.7}(b) as function of $Wi$. We observe that the spikes
in the local velocity signal occur about three time less frequently
than those in the injected power series. This results in poor statistics
and larger scatter, particularly for low values of $Wi$, but on the
other hand is consistent with the fact that the rare events in the
power time series are actually a result of space averaging (over the
entire volume of the flow cell) of individual rare events in local
velocity time series.

\begin{figure}[h]

\begin{centering}\includegraphics[width=12cm]{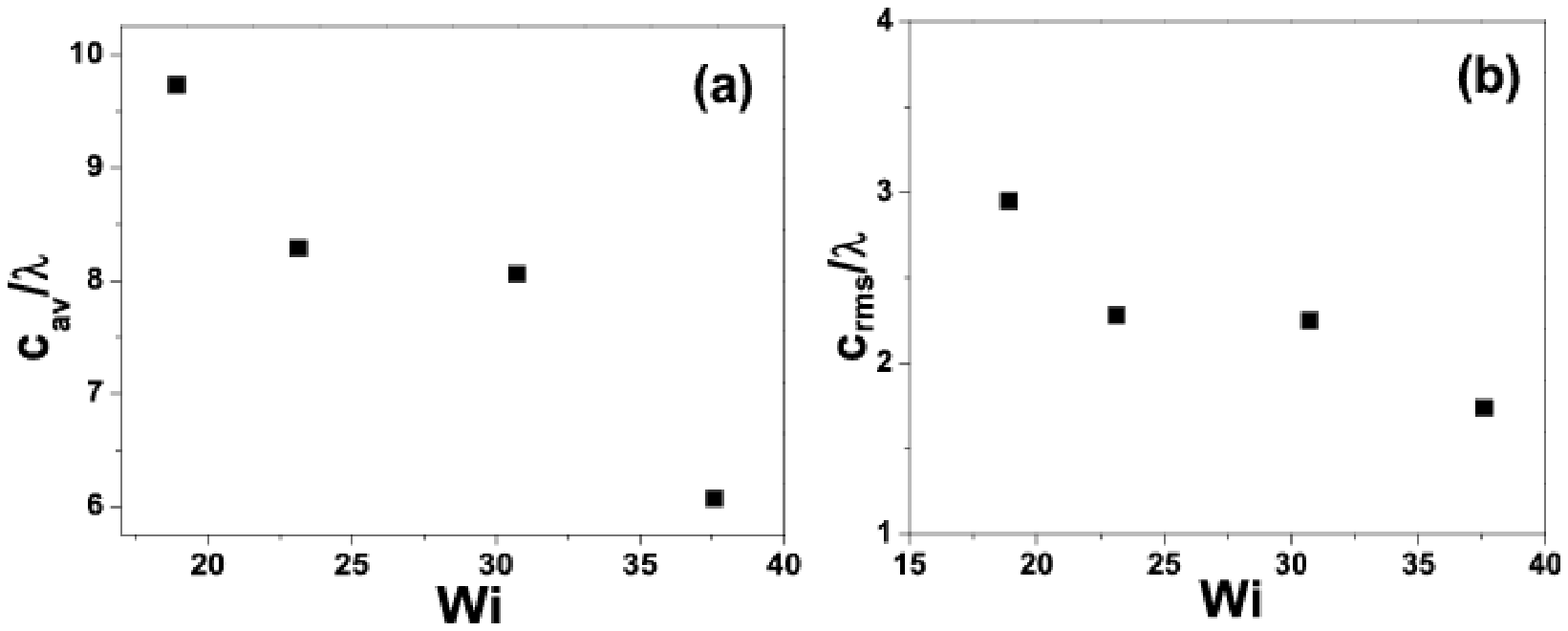} \end{centering}

\caption{(a) Normalized average spike width, $c_{av}/\lambda$, vs $Wi$.
(b) Normalized rms of the spike width, $c_{rms}/\lambda$, vs $Wi$.
Data were collected in constant $\Omega$ mode.\label{f.8} }
\end{figure}

We also computed the mean width of the spikes determined by the procedure
described above. The central part of each spike was fitted by a Gaussian
function, and all the fit widths $c$ for a given $Wi$ were averaged
together. The results for the spikes in the series of injected power
in the $\Omega$-mode are shown in Fig.~\ref{f.8}.

\subsection{Flow structure and statistical properties of the velocity field\label{sub:Flow-structure}}

PIV and LDV techniques have been used in the past in various experiments
and setups~\cite{NJP,taylorhypoth,thesis,my,teoprl} to characterize
the structure of the flow field, to obtain velocity profiles and to
infer and visualize the topology of the elastic von Karman flow. The
visual characterization of the flow structure in a regime of elastic
turbulence has been initially obtained from a few snapshots of the
flow, by seeding it with light reflecting flakes ($1\%$ of Kalliroscope
liquid)~\cite{NJP}. Already such qualitative visualization revealed
the presence of a big persistent toroidal vortex, which had the dimension
of the whole setup. The average flow velocity along the radial direction
was measured by LDV in a few points, and the results were quite well
compatible with the presence of the big vortical structure~\cite{NJP}.

\begin{figure}
\begin{centering}\centering \includegraphics[width=14cm]{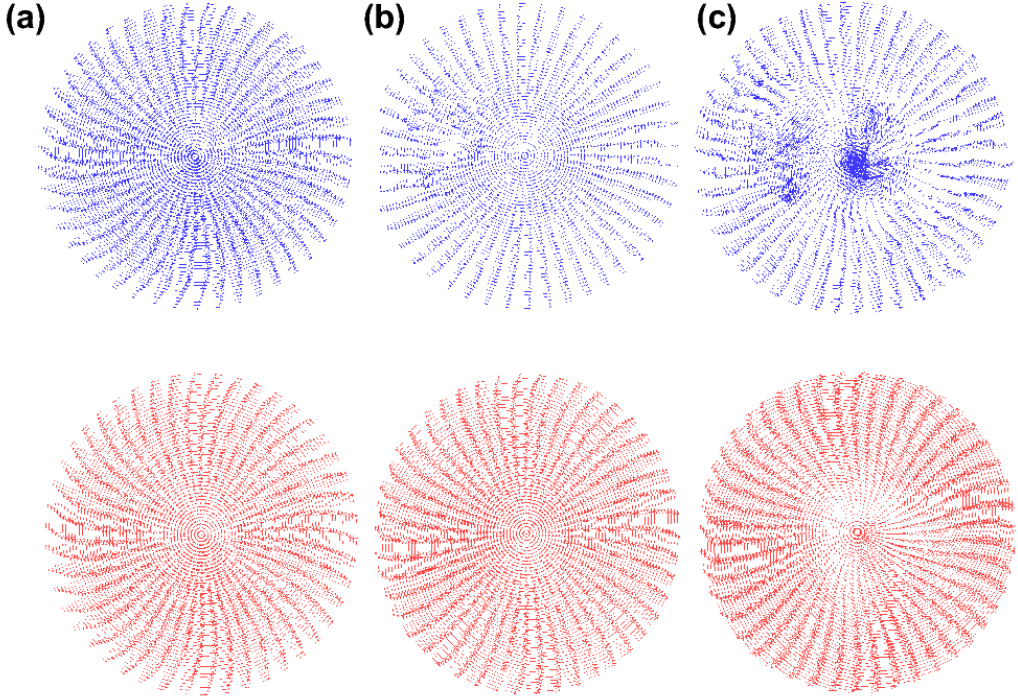} \end{centering}

\caption{Instantaneous (upper row) and time averaged (lower row) velocity
fields for different values of $Wi$: (a) $Wi=2.48$, (b) $Wi=9.88$,
(c) $Wi=18.96$. The data were collected in setup $1$ at middle distance
between plates.}

\label{f.9}
\end{figure}

\begin{figure}[h]

\begin{centering}\includegraphics[width=10cm]{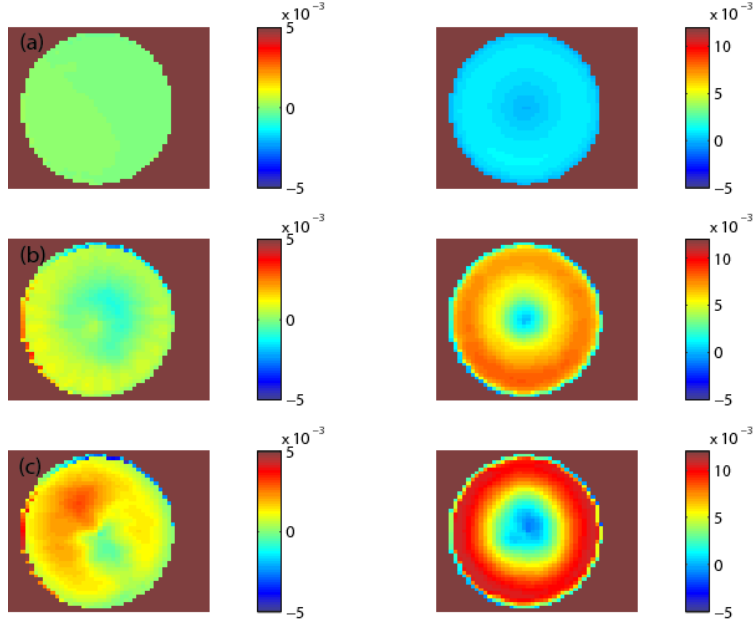} \end{centering}

\caption{Distribution of the radial velocity (left column) and the azimuthal
velocity components (right column) for different values of $Wi$:
\textbf{(a)} $Wi=2.48$, \textbf{(b)} $Wi=9.88$, \textbf{(c)} $Wi=18.96$.
The data were collected in setup $1$ at middle distance between plates.}

\label{f.10}
\end{figure}

\begin{figure}[h]

\begin{centering}\includegraphics[width=10cm]{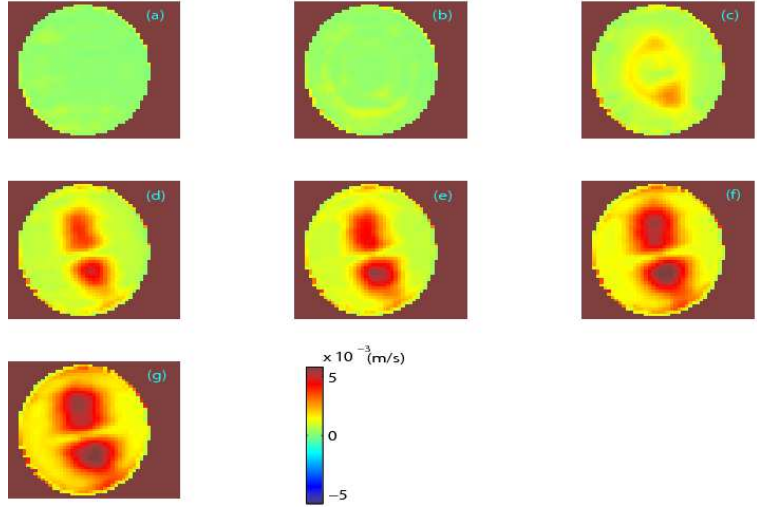} \end{centering}

\caption{Fields of fluctuations of the azimuthal velocity at different $Wi$:
\textbf{(a)} $Wi=8.32$, \textbf{(b)} $Wi=9.88$, \textbf{(c)} $Wi=11.1$,
\textbf{(d)} $Wi=12.72$, \textbf{(e)} $Wi=13.83$, \textbf{(f)} $Wi=17$,
\textbf{(g)} $Wi=19$. The data were collected in the setup $1$ at
middle distance between plates.}

\label{f.11}
\end{figure}

\begin{figure}[h]

\begin{centering}\includegraphics[width=12cm]{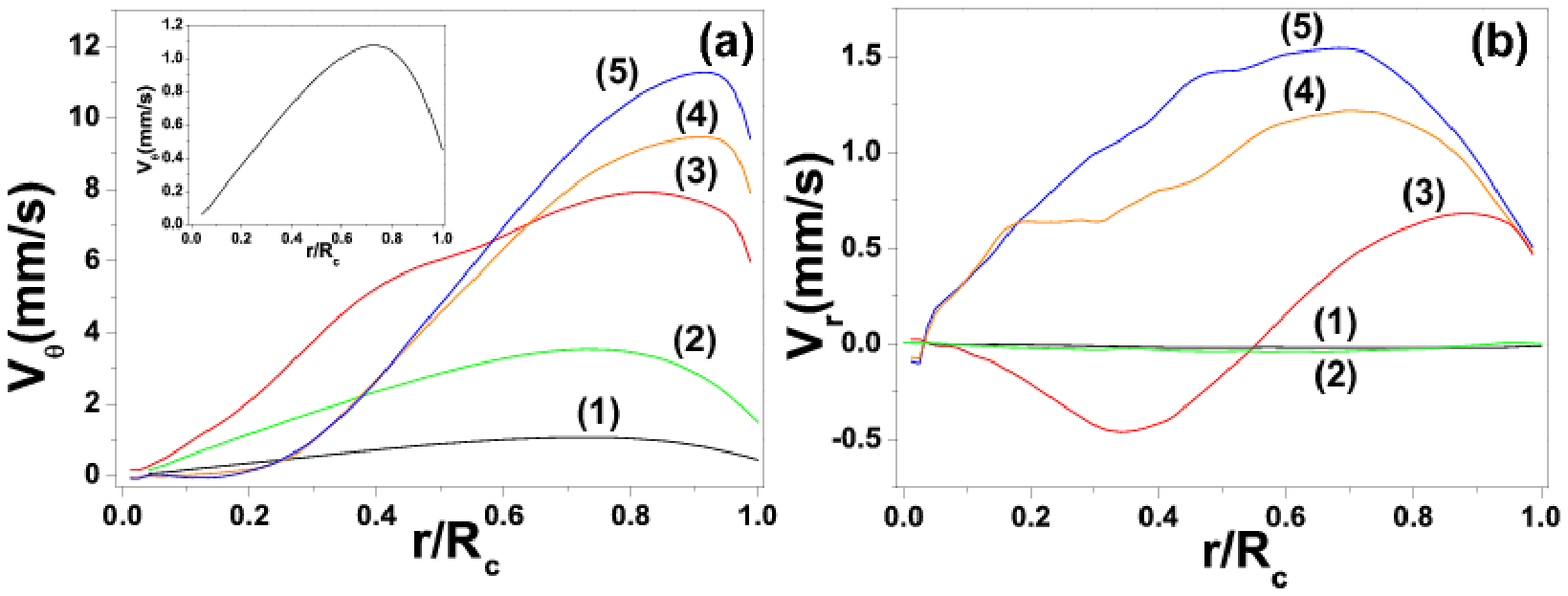} \end{centering}

\caption{\textbf{(a)} Profiles of the average azimuthal velocity, $V_{\theta}$,
for several values of $Wi$. The inset shows a typical laminar profile
of the azimuthal velocity component. \textbf{(b)}Profiles of the average
radial velocity, $V_{r}$, for several values of $Wi$. The labels
are: \textbf{(1)}-$Wi=2.48$, \textbf{(2)}-$Wi=4.41$, \textbf{(3)}-$Wi=11.1$,
\textbf{(4)}-$Wi=15$, \textbf{(5)}-$Wi=18$. Data were collected
in set-up $1$ at middle distance between plates.\label{f.12} }
\end{figure}

With PIV we obtained a more complete and quantitative characterization
of the global flow structure and of its dynamics. Some instantaneous
vector fields of the horizontal components of the velocity, as well
as the average of $2000$ instantaneous fields taken for several values
of $Wi$ at a middle distance between the plates in the setup $1$
are presented in Figs.~\ref{f.9} and~\ref{f.10}. The upper row
of Fig.~\ref{f.9} shows the instantaneous vector fields at three
increasing values of $Wi$, while the lower row presents averaged
vector field at the same $Wi$. One can easily identify the core of
the toroidal vortex that appears in all images above the threshold
of the instability, and a spiral vortex that additionally occurs in
the elastic turbulence regime. These flow structures and their reorganization
can be also observed in separate presentations of the averaged azimuthal
and radial components, and rms of fluctuations of the azimuthal velocity
component of the velocity field at the same $Wi$ (see in Fig.~\ref{f.10},\ref{f.11}).
Below the elastic instability the profile of the average azimuthal
velocity, $V_{\theta}$, exhibits a linear increase along the radius
with a slope $\Omega^{-1}$ which corresponds to a rigid body rotation
(see the inset in Fig.~\ref{f.12}a), while there is no motion in
the radial direction, $V_{r}\approx0$ (Fig.~\ref{f.12}b). Above
the onset of the instability one can clearly see the creation of the
core of the toroidal vortex at the cell center, and the restructuring
of the radial motion~\cite{thesis,my,teoprl}.

Large toroidal vortexes driven by the hoop stress are actually quite
well known to appear in swirling flows of elastic fluids~\cite{bird,Boger2}.
The inhomogeneity of the shear rate profile in the primary laminar
flow has long been recognized as their common origin. In our system
this vortex first arises as a stationary structure at low shear rates,
and leads to some increase in flow resistance even before the elastic
instability~\cite{NJP}. Therefore, one concludes that the transition
to elastic turbulence in the swirling flow between two plates is mediated
by this vortex. The toroidal vortex is providing a smooth, large scale
velocity field (see lower panels in Fig.~\ref{f.9}), which is randomly
fluctuating in time (upper panels in Fig.~\ref{f.9}), and in which
the fluid and the embedded stress tensor are chaotically advected.
This type of advection can create variations of stress in a range
of smaller scales, which may cause small scale fluid motion~\cite{Volodya,Volodya2}.
This is analogous to the generation of small scale variations of scalar
concentration in chaotic mixing by large fluctuating vortexes.

\begin{figure}[h]
\begin{centering}\centering \includegraphics[width=12cm]{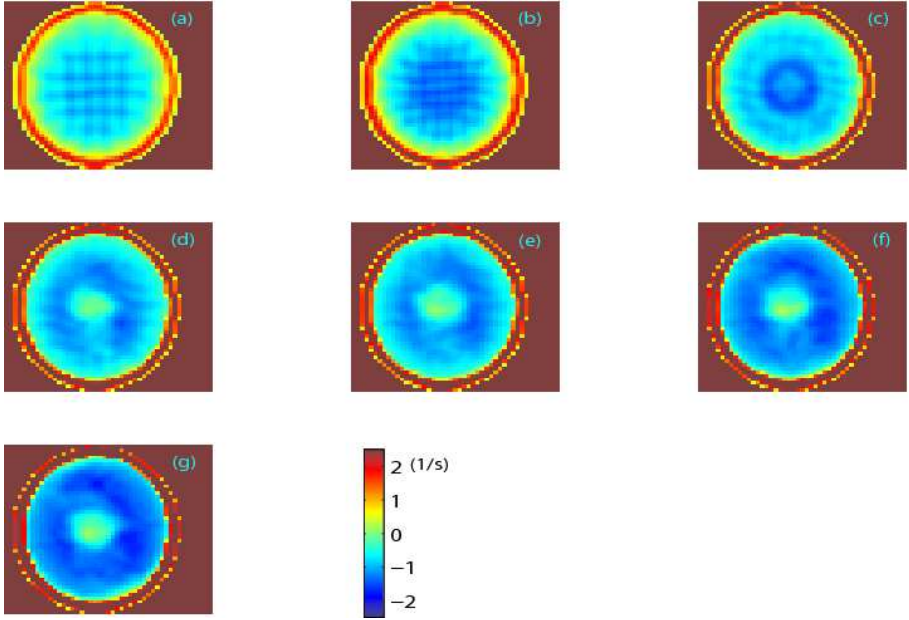} \end{centering}

\caption{Average vertical component of the vorticity, $\omega_{z}$, at different
$Wi$: \textbf{(a)} $Wi=8.32$, \textbf{(b)} $Wi=9.88$, \textbf{(c)}
$Wi=11.1$, \textbf{(d)} $Wi=12.72$, \textbf{(e)} $Wi=13.83$, \textbf{(f)}
$Wi=17$, \textbf{(g)} $Wi=19$. Data were collected in setup $1$
at middle distance between plates. The gridded pattern visible in
panels (a), (b) is an artifact of combined peak locking and numerical
differentiation effects and should be disregarded.}

\label{f.13}
\end{figure}

Above the instability threshold, the large scale toroidal vortex is
also responsible for the ring-shaped topology of the fluctuation field
of the azimuthal velocity, $V_{\theta}^{rms}\equiv\overline{V_{\theta}^{2}}^{1/2}$
(Fig.~\ref{f.11}). The velocity fluctuations in the laminar regime
are only due to instrumental noise (panel (a) in Fig.~\ref{f.11}).
The average radial velocity changes its sign at about a half of the
radius (see Fig.~\ref{f.12}b). At higher values of $Wi$, the toroidal
vortex forces the transition to elastic turbulence. This is accompanied
by a second reorganization of the flow structure, with the appearance
of a fluctuating spiral vortex (Figs.~\ref{f.9},\ref{f.10}) and
takes part in the scaling regime of $P$, $P_{rms}$ (see Fig.~\ref{f.3}).
Movie animations of the velocity fields evidence that this secondary
fluctuating vortex is carried around and {}``rides'' on top of the
primary vortical flow. At the same time the circular symmetry of $V_{\theta}^{rms}$
is broken into a dipolar one (see Fig.~\ref{f.11}). The radial profiles
of $V_{\theta}$ and $V_{r}$ also change drastically. The structural
transitions in the flow are also reflected in the average radial gradients
of the azimuthal velocity and in the average vorticity field, as presented
in Fig.~\ref{f.13} for various values of $Wi$.

\begin{figure}[h]
\begin{centering}\centering \includegraphics[width=9cm]{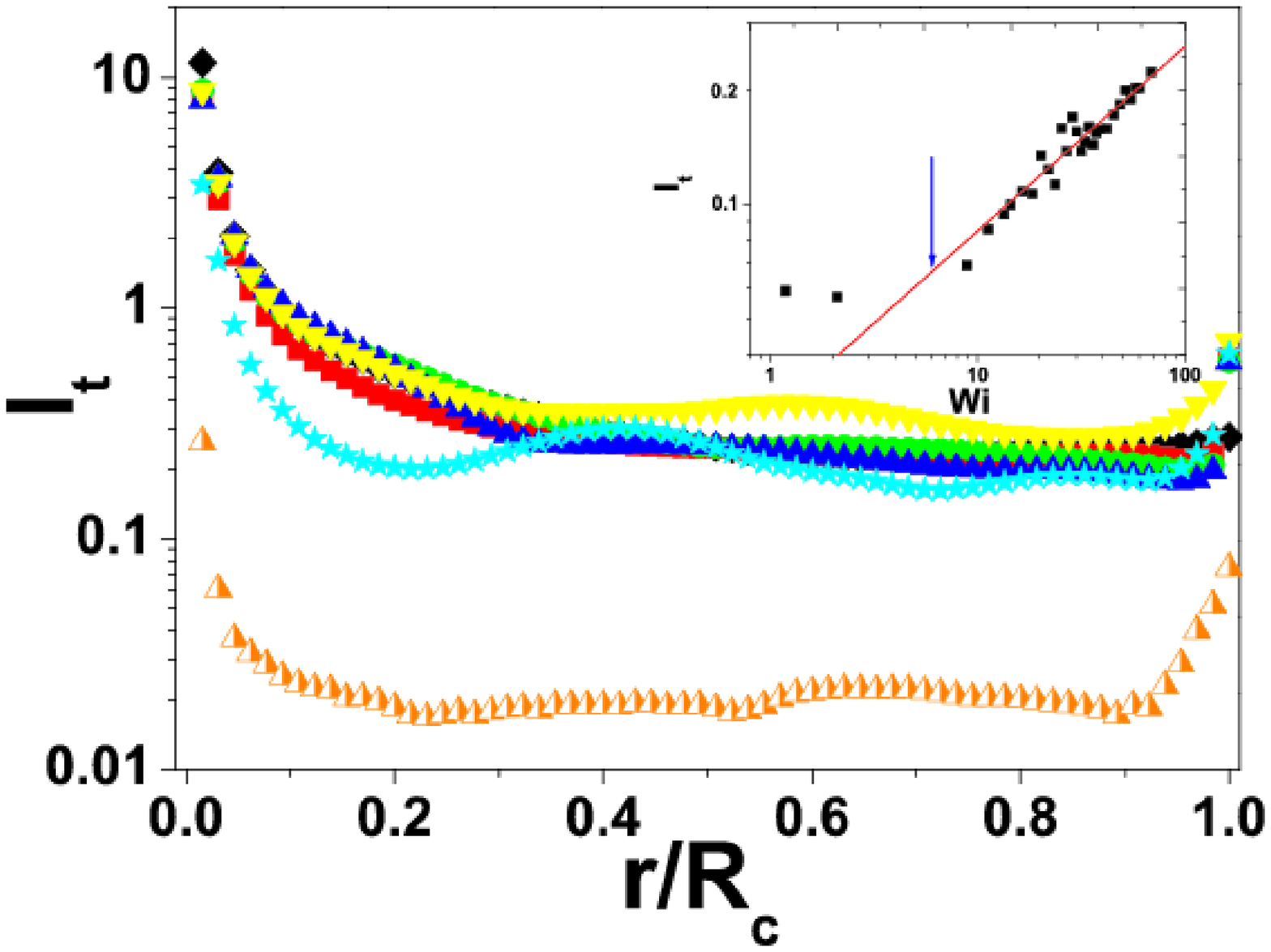} \end{centering}

\caption{Turbulence intensity versus the reduced radial coordinate at different
$Wi$: diamonds-$Wi=31.56$, squares- $Wi=27.73$, circles-$Wi=23.67$,
up triangles-$Wi=19.29$, down triangles-$Wi=15.48$, stars-$Wi=7.54$,
half filled triangles-$Wi=2.82$. The inset shows the dependence of
the turbulence intensity on $Wi$ at $r/R_{c}=0.85$ obtained from
LDV measurements. The full line is a power law fit, $I_{t}\propto Wi^{0.49\pm0.06}$.
The arrow indicates the onset of the elastic instability. The data
were collected in setup 2.\label{f.14} }
\end{figure}

Another way to characterize a turbulent flow is to display the radial
profiles (averaged over the azimuthal coordinate) of the turbulent
intensity, defined as $I_{t}=\frac{V_{\theta}^{rms}}{V_{\theta}}$
at several values of $Wi$ (see Fig.~\ref{f.14})~\cite{teoprl}.
The velocity fluctuations in the laminar regime occur only due to
instrumental errors. Above the elastic transition $I_{t}$ increases
sharply but remains rather uniform on the level between $20$ and
$30\%$ in a peripheral region for the radius ratios, $r/R_{C}$,
between $0.2$ and $1$. In the regime of elastic turbulence $I_{t}$
further increases, and its dependence on $Wi$, presented in the inset
in Fig.~\ref{f.14}, exhibits the power-law scaling, $I_{t}\sim Wi^{0.49\pm0.06}$~\cite{thesis,my}.

\begin{figure}
\centering \includegraphics[width=14cm]{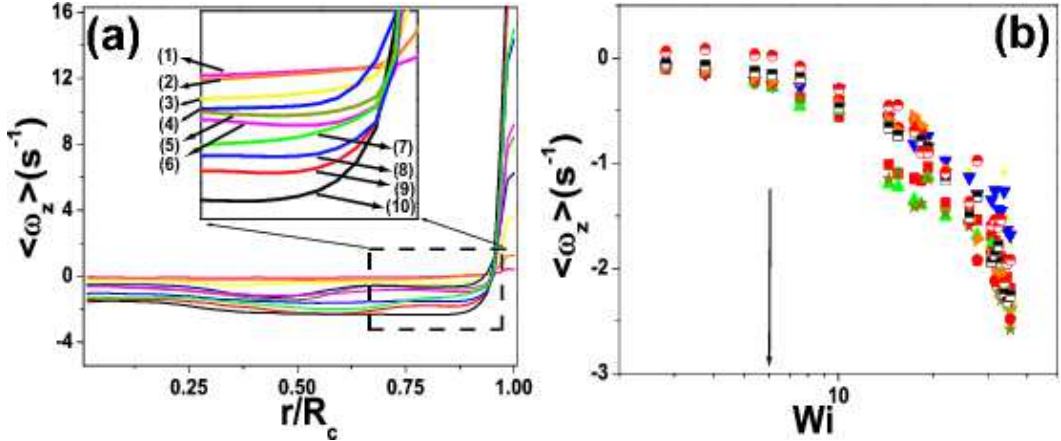}

\caption{\textbf{(a)} Profiles of the vorticity $\langle\omega_{z}\rangle$
averaged spatially in the azimuthal direction and temporally over
$2000$ images, for different $Wi$. The labels are: \textbf{(1)}-$Wi=2.82$,
\textbf{(2)}-$Wi=5.4$, \textbf{(3)}-$Wi=10$, \textbf{(4)}-$Wi=14.48$,
\textbf{(5)}-$Wi=17.43$, \textbf{(6)}-$Wi=19.29$, \textbf{(7)}-$Wi=26.13$,
\textbf{(8)}-$Wi=27.73$, \textbf{(9)}-$Wi=32.3$, \textbf{(10)}-$Wi=34.5$.
\textbf{(b)} Dependence of the average vorticity on $Wi$ at different
radial positions $r/R_{c}$,: left triangles-$0.1$, down triangles-$0.2$,
squares-$0.33$, up triangles-$0.4$, stars-$0.5$, circles-$0.66$,
right triangles-$0.7$, half filled squares-$0.8$, half filled circles-$0.9$.
The arrow marks the onset of the elastic instability. The data were
collected in setup $2$.\label{f.15} }
\end{figure}

\begin{figure}
\centering \includegraphics[width=14cm]{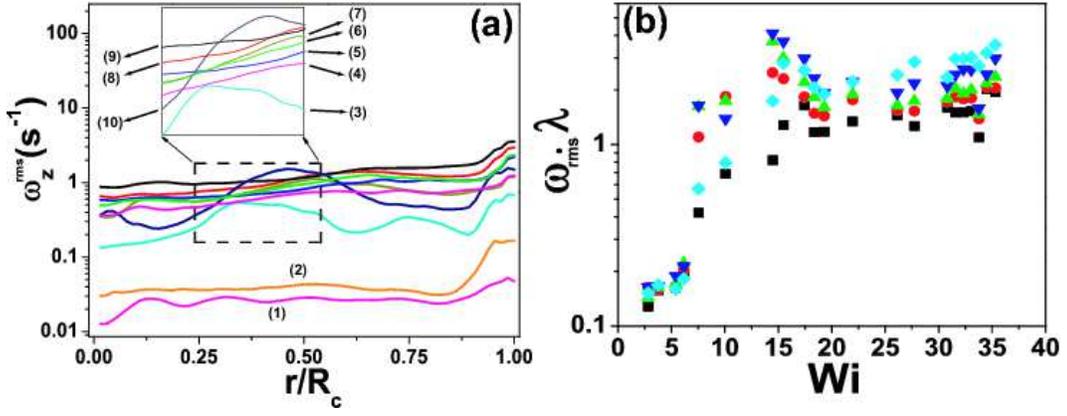}

\caption{\textbf{(a)} Profiles of the rms of the fluctuations of the vorticity,
$\omega_{z}^{rms}$, at different $Wi$. The labels are: \textbf{(1)}-$Wi=2.82$,
\textbf{(2)}-$Wi=5.4$, \textbf{(3)}-$Wi=10$, \textbf{(4)}-$Wi=14.8$,
\textbf{(5)}-$Wi=17.43$, \textbf{(6)}-$Wi=19.29$, \textbf{(7)}-$Wi=26.3$,
\textbf{(8)}-$Wi=26.3$, \textbf{(9)}-$Wi=27.73$, \textbf{(10)}-$Wi=32.3$,
\textbf{(b)} Dependence of the scaled rms of the vorticity, $\omega_{rms}\cdot\lambda$,
on $Wi$ at different radial positions $r/R_{c}$: full squares, $0.2$;
open squares, $0.33$; full circles, $0.4$; open circles, $0.5$;
diamonds, $0.66$. The data were collected in setup $2$.\label{f.16} }
\end{figure}

The PIV measurement of time dependent velocity fields allowed us to
calculate the average velocity gradients and vertical vorticity, and
rms of their respective fluctuations, without involving the Taylor
hypothesis that can be questionable for a smooth random flow~\cite{taylorhypoth,thesis}.
The typical radial distribution of the velocity gradients and of vorticity,
averaged spatially over the azimuthal coordinate and temporally over
$2000$ images is rather uniform in the bulk for all $Wi$, but increases
sharply near the wall (see Fig.~\ref{f.15}(a)), while the rms of
the velocity gradients and vorticity gradually increases with the
radius (Fig.~\ref{f.16}(a)). The dependence of the average vertical
vorticity on $Wi$ is displayed in Fig.~\ref{f.15}(b). At higher
values of $Wi$ one finds a gradual growth of the vorticity both in
the bulk and near the wall. On the other hand, the plot in Fig.~\ref{f.16}(b)
shows the $Wi$ dependence of the rms of the vorticity, $\omega_{rms}$,
scaled by $\lambda$, at several locations along a radius in the bulk.
The scaled rms of the vorticity saturates in the elastic turbulence
regime at a $Wi$ between $12$ and $35$, that is consistent with
the recent theoretical prediction. Similar behavior is shown by all
the components of the velocity gradient, and is consistent with the
theoretical predictions, though the saturation value $\omega_{z}^{rms}\cdot\lambda\approx2$
is higher than expected~\cite{Volodya,Volodya2}. The latter means
probably that the nonlinearity of the polymer elasticity also contributes
to the saturation of the elastic stress~\cite{lebedev,chertkov}
(see further discussion of this issue).

\begin{figure}
\centering \includegraphics[width=16cm]{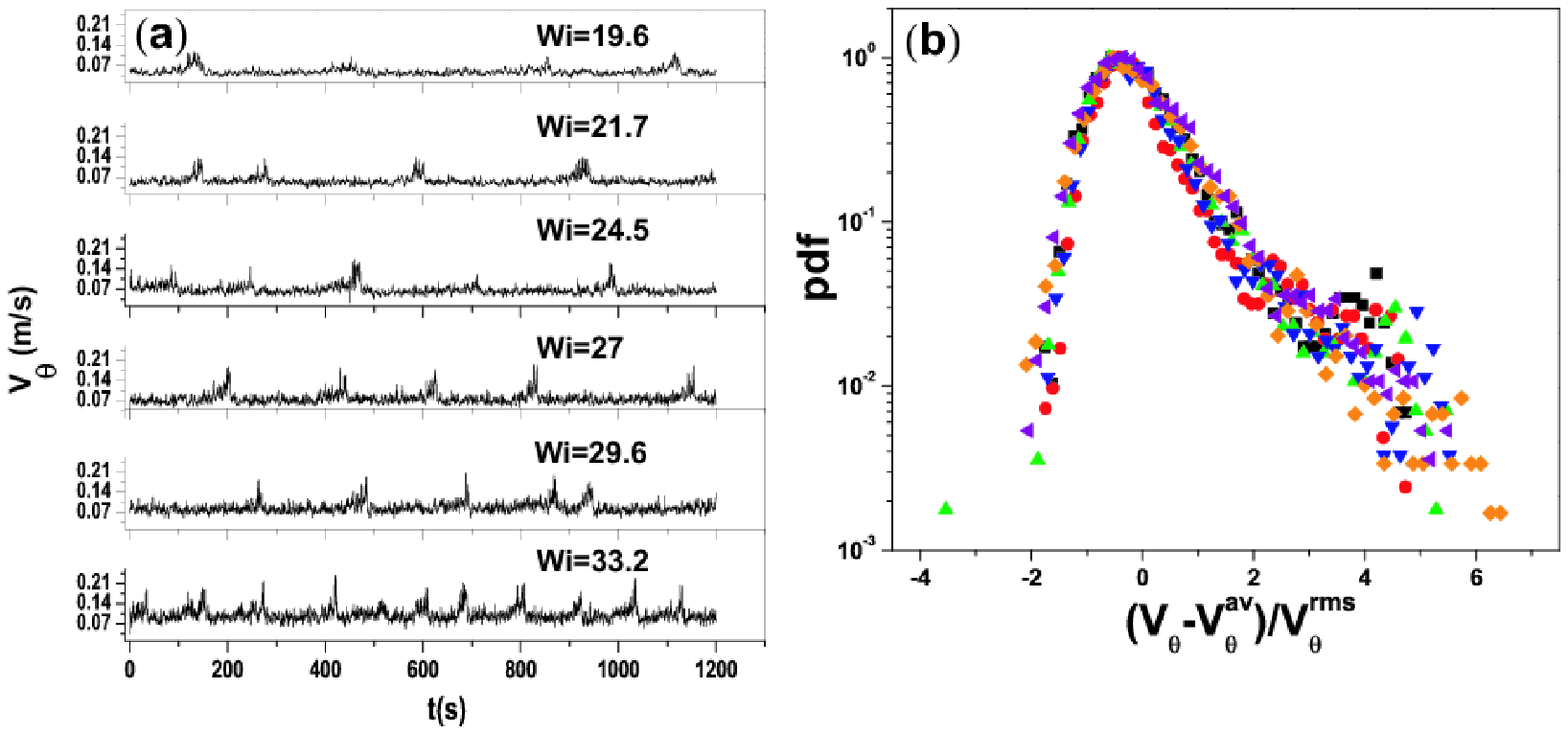}

\caption{(a) Time series of azimuthal velocities, $V_{\theta}$, for several
values of $Wi$. (b) PDFs of normalized azimuthal velocities for various
$Wi$: squares-$19.6$, circles-$21.7$, up triangles-$24.5$, down
triangles-$27$, diamonds-$29.6$, left triangles-$33.2$. Data were
collected in setup $2$, at $z=d/2$ and $r=2R_{c}/3$.\label{velstat} }
\end{figure}

Time series of azimuthal velocity obtained from LDV measurements for
several values of $Wi$ are displayed in Fig.~\ref{velstat}(a).
As $Wi$ increases, the velocity signal becomes increasingly intermittent.
PDFs of the normalized azimuthal velocity, $(V_{\theta}-V_{\theta}^{av})/V_{\theta}^{rms}$,
are shown in Fig.~\ref{velstat}(b). The right side skewness of the
PDFs is due to the rare events (spikes) which intermittently occur
in the velocity signal (Fig.~\ref{velstat}(a)), which are probably
related to the dynamics of the randomly fluctuating spiral vortex
previously discussed.

\begin{figure}
\centering \includegraphics[width=17cm]{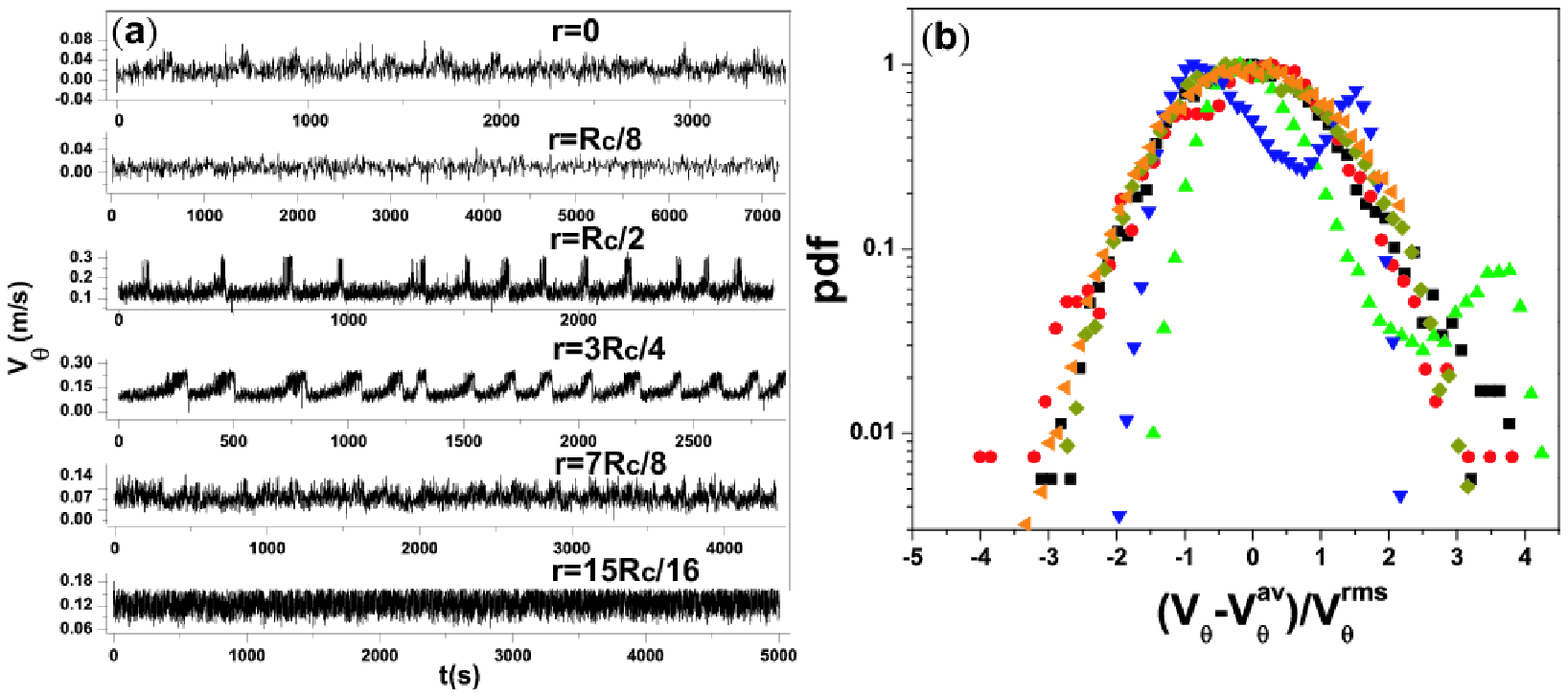}

\caption{(a) Time series of azimuthal velocities, $V_{\theta}$, for $Wi=36.1$
at several radial positions, $r$. (b) PDFs of the normalized azimuthal
velocity, for $Wi=36.1$, at several radial positions $r$: squares-$0$,circles-$R_{c}/8$,up
triangles-$R_{c}/2$,down triangles-$3R_{c}/4$,diamonds-$7R_{c}/8$,
left triangles-$15R_{c}/16$. The data were collected in setup $2$,
at $z=d/2$.\label{radvelstat} }
\end{figure}

In order to understand better the structure of the flow in the regime
of elastic turbulence, we have studied the dependence of the statistics
of the azimuthal velocity on the radial coordinate, for a fixed value
of $Wi=36.1$ and a given vertical coordinate, $z=d/2$. As illustrated
in Fig.~\ref{radvelstat}(a), in the central ($r=0$) and peripheral
($r=15R_{c}/16$) flow regions, the velocity signal shows no significant
intermittency, whereas around $r=R_{c}/2$ it is strongly intermittent.
The radial dependence of the level of intermittency reflects the presence
of the randomly fluctuating spiral vortex (where, probably, $r=R_{c}/2$
corresponds to the arm of the spiral). This behavior is also reflected
in the PDFs of the normalized azimuthal velocity presented in Fig.~\ref{radvelstat}(b):
near the center and the vertical wall of the cell, the distributions
are symmetric and single peaked whereas around $r=R_{c}/2$ they become
strongly skewed and doubly peaked.

\begin{figure}
\centering \includegraphics[width=8cm]{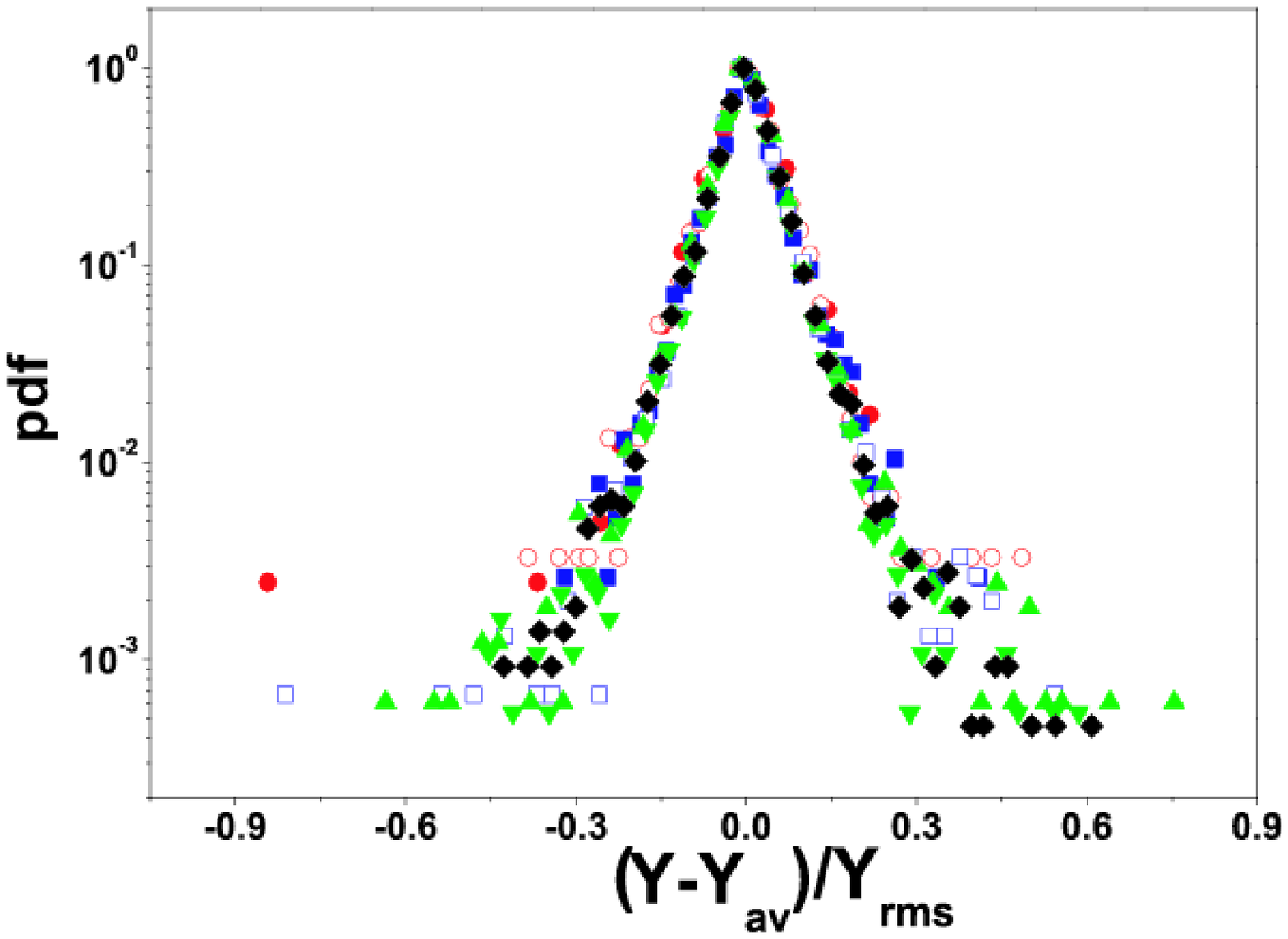}

\caption{PDFs of normalized accelerations for various $Wi$: full circles-$14$;
open circles-$16.2$; full squares-$19.8$; open squares-$22.3$;
up-triangles-$26.3$; down-triangles-$30.7$; diamonds-$38.9$. The
data were collected in setup $2$ at $z=d/2$ and $r=2R_{c}/3$.}

\label{f.19}
\end{figure}

Fig.~\ref{f.19} shows PDFs of the normalized accelerations as calculated
from the temporal LDV measurements, $(Y(t)-Y_{av})/Y_{rms}$, where
$Y(t)\equiv\frac{dV(t)}{dt}V_{av}^{-1}$. We note that all the data
for several values of $Wi$ collapse on a single curve remarkably
well, and that moreover the PDFs show clear exponential tails.

The PIV data allow us also to calculate various correlation and structure
functions of velocity, velocity gradients, and vorticity. Their scalings
can give information on the degree of deviation from a Gaussian random
field, a standard way in hydrodynamics to quantify intermittency and
to compare it with the corresponding scaling in known cases.

\begin{figure}[h]
\begin{centering}\includegraphics[width=8cm]{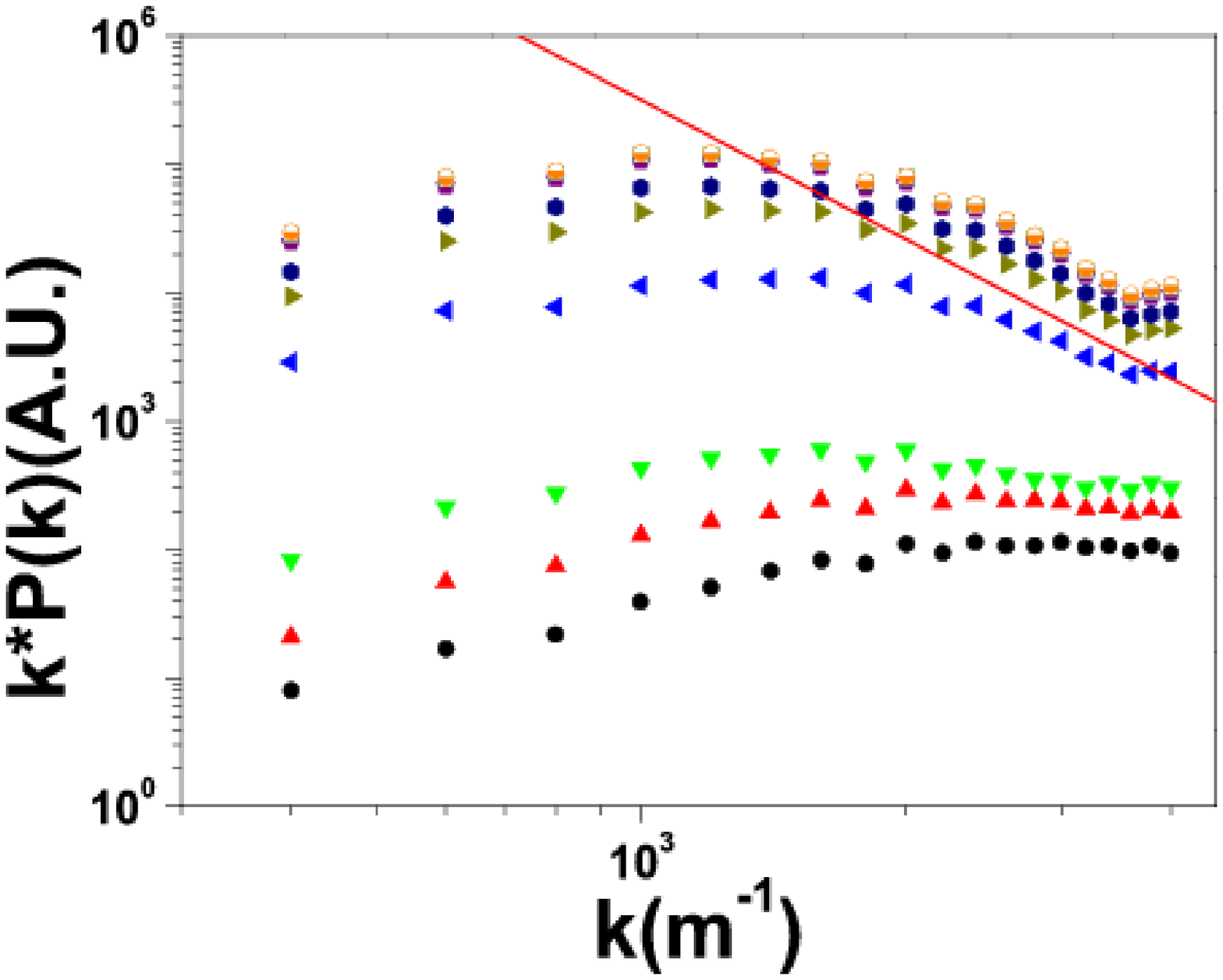} \end{centering}

\caption{Power spectra of the fluctuations of the azimuthal velocity component
at different $Wi$: circles-$Wi=4.41$, up triangles-$Wi=5.72$, down
triangles-$Wi=8.32$, left triangles $Wi=11.1$, right triangles-$Wi=12.7$,
hexagons-$Wi=13.8$, diamonds-$Wi=16$, half filled squares-$Wi=18$,
empty circles-$Wi=19$. The solid line is the power law $k^{-3.5}$.
The data were collected in setup $1$.\label{f.20} }
\end{figure}

The spatial power spectra of the velocity were obtained from PIV data
by averaging $2000$ instantaneous spatial spectra. Although the finite
spatial resolution of PIV, which limits the accessible range of wave
numbers, leads to an artificial cut-off at $k\sim3000m^{-1}$, a power
law decay $k^{-\delta}$ with $\delta\simeq3.5$ is clearly observed
(Fig.~\ref{f.20})~\cite{thesis,my}. This decay is consistent with
the scaling obtained earlier for the velocity power spectra in the
frequency domain~\cite{Nat,NJP,my}. The large value of the exponent
$\delta$ implies that the power of fluctuations decays very quickly
as the size of eddies decreases. The main contribution to the fluctuations
of both the velocity and the velocity gradients is therefore due to
the largest eddies, and the power of the latter should scale as $k^{-(\delta-2)}$.

The agreement between the velocity spectra measured at a single point
in frequency domain~\cite{Nat,NJP} and the spectra measured directly
in $k$-domain~\cite{thesis,my} deserves a brief discussion. Following
Lumley~\cite{lumley}, the relation between the spatial spectrum
$P(k)$ and the frequency spectrum $P(f)$ can be written as \begin{equation}
P(k)=V\cdot P(f)-\frac{I_{t}^{2}}{2}\cdot\frac{d^{2}\left(k^{2}P(k)\right)}{dk^{2}}+O\left(I_{t}^{4}\right),\label{eq:spectra}\end{equation}
 where $I_{t}=V_{\theta}^{rms}/V_{\theta}$, $V_{\theta}^{rms}\equiv{\overline{V_{\theta}^{2}}}_{t}^{1/2}$
and $V_{\theta}$ are the rms of fluctuations of the azimuthal velocity
and the average azimuthal velocity, respectively. If $P(k)\propto k^{\delta_{2}}$
and $P(f)\propto f^{\delta_{1}}$, the equation above leads to: \begin{equation}
{\delta_{2}-\delta_{1}}\propto\frac{\log\left[1+\frac{I_{t}^{2}}{2}\cdot\delta_{2}\cdot(\delta_{2}+2)\right]}{\log(k)}\,.\label{eq:powerlawexponents}\end{equation}
 If one plugs into the last equation $\delta_{2}\approx-3.5$, the
difference between the exponents is (for $k\approx1000m^{-1}$) as
small as ${\delta_{2}-\delta_{1}}\approx0.1$. Thus, the experimental
resolution does not allow us to observe the difference in the scaling
exponents for the spatial and temporal spectra~\cite{thesis}.

\begin{figure}
\centering \includegraphics[width=8cm]{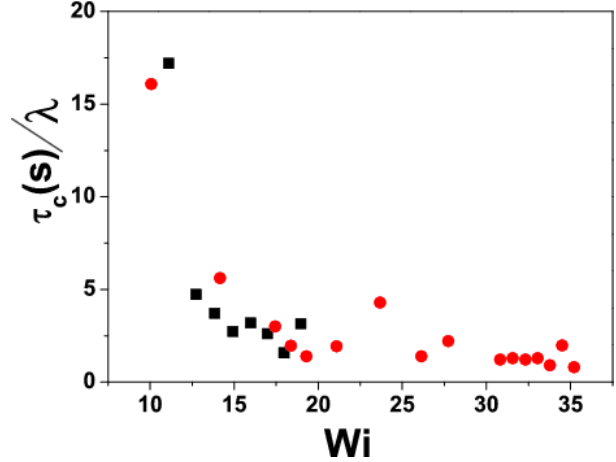}

\caption{Scaled Eulerian correlation times of the azimuthal velocity component
as function of $Wi$: squares-setup $1$, circles-setup $2$.\label{f.21} }
\end{figure}

Next, let us focus on the typical spatial and temporal correlation
scales in elastic turbulence. A typical scale, at which elastic stress
is correlated can be estimated as $L=2\pi\int P(k)dk/\int k\, P(k)dk$.
In our setup and in the elastic turbulence regime, one gets $L\simeq5.9$
cm, i.e.\ on the order of the cell size. The Eulerian correlation
time is defined as $\tau_{c}=\int t\, C(t)dt/\int C(t)dt$, where
$C(t)=\overline{V(T)V(T+t)}/\left(V^{rms}\right)^{2}$ is the Eulerian
correlation function. Figure~\ref{f.21} presents the normalized
$\tau_{c}/\lambda$ as a function of $Wi$ at some radial positions
in the cell. The correlation time drops significantly in the transition
region and then saturates in the elastic turbulence regime, as the
rms of the vorticity (and velocity gradients) fluctuations do. The
related saturation of both the Lyapunov exponents of the Lagrangian
trajectories, already reported~\cite{teo}, and the reduced velocity
correlation time, $\tau_{c}/\lambda$, shown in Fig.~\ref{f.21}
(in the same range of $Wi$) just reinforces the observation. We point
out that the inverse saturated value of correlation time is of the
order of the saturated value of $\omega_{z}^{rms}$ and $\lambda^{-1}$,
as predicted by theory~\cite{Volodya,Volodya2}.

\begin{figure}
\begin{centering}\centering \includegraphics[width=8cm]{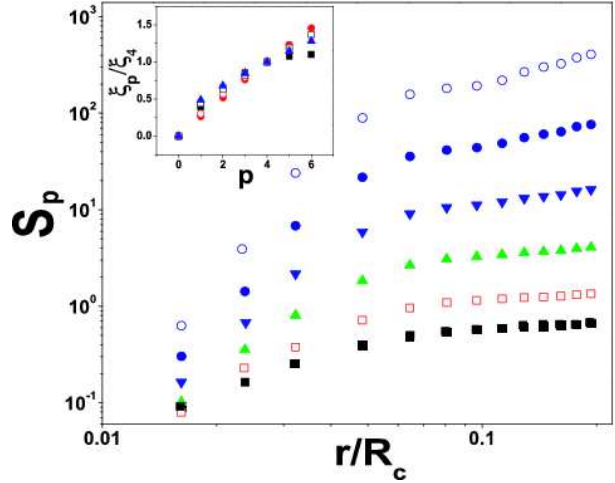} \end{centering}

\caption{Structure functions of the velocity gradients of order $p$ up to
$6$ as a function of the scaled radius. Inset: normalized scaling
exponent vs.\ $p$ for: injected power $P_{\Omega}$- full circles;
injected power $P_{T}$- open circles; passive scalar- full squares;
vorticity- open squares; velocity gradient- up-triangles.\label{f.22} }
\end{figure}

Using the PIV data one can also calculate the structure functions
of the velocity gradients, defined as \begin{equation}
S_{p}(r)=\left\langle \left|\frac{\partial V_{\theta}(r_{0}+r)}{\partial r}-\frac{\partial V_{\theta}(r_{0})}{\partial r}\right|^{p}\right\rangle _{r_{0}}\label{eq:structureFunction}\end{equation}
 and shown in Fig.~\ref{f.22}. As for inertial turbulence, we can
look for a scaling in the spatial range corresponding to the power
law decay of the velocity spectra (Fig.~\ref{f.20}) in the form:
$S_{p}(r)=r^{\zeta_{p}}$. The dependence of the normalized scaling
exponent, $\zeta_{p}/\zeta_{4}$, on the order of the structure functions,
$p$, is presented in the inset of Fig.~\ref{f.22} for all flow
parameters mentioned above. For comparison, the structure functions
of the vorticity, of the injected power at constant $\Omega$, $P_{\Omega}$,
and constant torque, $P_{T}$, were also calculated, and the results
of the analysis for the normalized scaling exponent were also plotted
in the inset of Fig.~\ref{f.22}. All these dependencies are rather
close to that obtained for a passive scalar in elastic turbulence,
presented on the same plot and based on our published data~\cite{Teo,thesis};
all of them deviate from a linear dependence that is characteristic
of a Gaussian distribution~\cite{Shr}.

It is worth to make here some remarks about the applicability of the
Taylor hypothesis to flows which are spatially smooth random in time,
such as is elastic turbulence, which we have studied experimentally
in more detail in~\cite{taylorhypoth}. On one hand, the comparison
of the power spectra of the velocity in frequency and wave number
domain suggests that the Taylor hypothesis is applicable at least
in regions close to the wall. On the other hand, from Fig.~\ref{f.14}
one can learn that in the central flow region, at $r/R_{c}<0.2$,
the rms of the velocity fluctuations becomes much larger that the
mean velocity, fact which is normally thought to violate the assumptions
of the hypothesis. By detailed analysis of cross-correlations as well
as of the structure functions of the velocity fluctuations, it was
indeed shown in~\cite{taylorhypoth} that strong velocity fluctuations
near the cell center prevent, and flow smoothness and lack of scale
separations close to the boundaries limit the quantitative applicability
of the Taylor hypothesis. However, this deficiency can be corrected
by a proper choice of the advection velocity, so that a corrected
Taylor hypothesis is applicable to most parts of this chaotic flow.

\subsection{Velocity and velocity gradient profiles and the boundary layer problem\label{sub:Velocity-and-boundary}}

\begin{figure}[h]
\begin{centering}\centering \includegraphics[width=9cm]{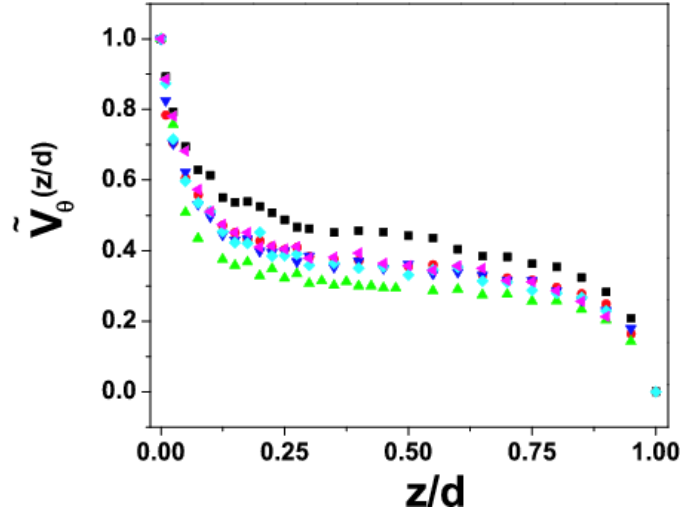} \end{centering}

\caption{Vertical profile of the average azimuthal velocity, normalized by
its maximum value at the disk, measured by LDV in setup~$2$, with
$d=2$cm, in a vertical plane at $r=3R_{c}/4$, in a 65\% saccharose
polymer solution, for several values of $\Omega$: full squares (black)-$5s^{-1}$,
circles (red)-$4s^{-1}$, up-triangles (green)-$3s^{-1}$, down-triangles
(blue)-$2.5s^{-1}$, diamonds (cyan)-$2s^{-1}$, left-triangles (magenta)-$1.5s^{-1}$.\label{f.23} }
\end{figure}

\begin{figure}[h]

\begin{centering}\centering \includegraphics[width=9cm]{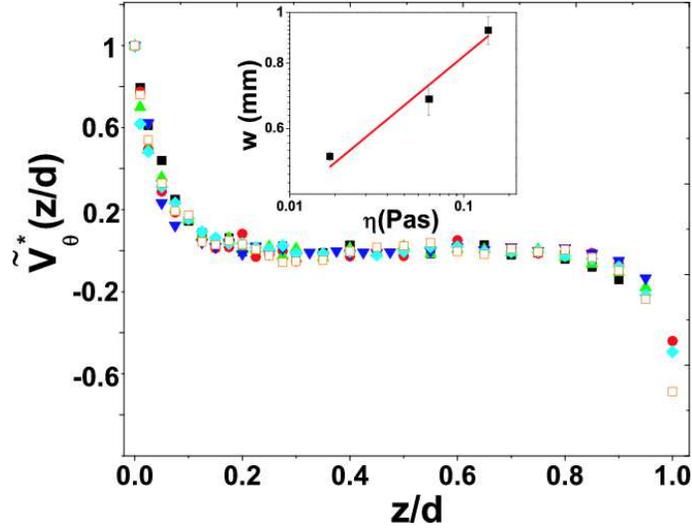} \end{centering}

\caption{Compensated vertical profile of the average azimuthal velocity $\tilde{V_{\theta}^{*}}$(see
text), for several values of $\Omega$: full squares-$5s^{-1}$; circles-$4s^{-1}$;
up-triangles-$3s^{-1}$; down-triangles-$2.5s^{-1}$; diamonds-$2s^{-1}$;
open squares-$1.5s^{-1}$. Inset: boundary layer thickness $w$ vs.\ solution
viscosity; the solid line is the fit $w\propto\eta^{0.26\pm0.05}$.
Data measured in the same setup as data of Fig.~\ref{f.23}. \label{f.24} }
\end{figure}

\begin{figure}
\includegraphics[width=9cm]{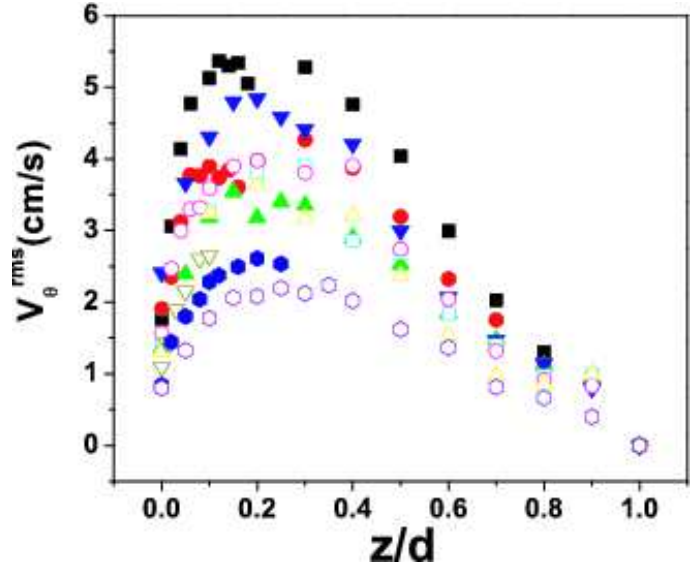}

\caption{Profile of the rms of the azimuthal velocity measured by LDV in a
vertical plane at $r=3R_{c}/4$, in setup $2$ with $d=1$cm height,
in 65\% of saccharose solution, for several values of $\Omega$: full
squares-$6rad/s$, full circles-$5.5rad/s$, full up triangles- $5rad/s$,
full down triangles-$5.5rad/s$, empty squares-$4rad/s$, empty circles-$3.7rad/s$,
empty up triangles-$3.5rad/s$, empty down triangles- $3rad/s$, full
hexagons-$2.5rad/s$, empty hexagons-$2rad/s$.\label{f.25} }
\end{figure}

To learn more about the flow field generated by elastic turbulence,
we investigated in detail the structure of horizontal and vertical
velocity boundary layers for different values of $Wi$ and fluid viscosity.
The vertical profile of the azimuthal velocity, $V_{\theta}$, in
the laminar flow of the pure solvent is nearly a straight line~\cite{NJP,my}.
The transition to elastic turbulence flow significantly changes the
distribution of $V_{\theta}^{av}$, producing a high shear layer near
the upper plate and a low shear region near the middle of the gap
(at $z/d=0.5$). Such a distribution of $V_{\theta}^{av}$ is reminiscent
of average velocity profiles in usual high $Re$ turbulence. In Fig.~\ref{f.23}
we present the average azimuthal velocity profiles measured by LDV
in a vertical plane at $r=3R_{c}/4$ of the setup $2$ with $d=2$cm
height in $65\%$ of saccharose solution and normalized by the velocity
value at the disk, $\tilde{V}_{\theta}(z/d)=V_{\theta}^{av}(z/d)/V_{\theta}^{up}$.
In Fig.~\ref{f.24} the same data for different $\Omega$ are collapsed
on a single curve, after subtraction the $z$-dependence of the average
velocity in the bulk for each curve, that is $\tilde{V}_{\theta}^{*}(z/d)=\tilde{V}_{\theta}(z/d)-\tilde{V}_{\theta}(1/2)-\frac{z}{d}\left.\frac{\partial\tilde{V}_{\theta}(z/d)}{\partial\left(z/d\right)}\right|_{1/2}$,
where $\frac{\partial\tilde{V}_{\theta}(z/d)}{\partial\left(z/d\right)}$
is obtained from a linear fit of the data points. Similar velocity
profiles were also obtained with polymer solutions with saccharose
content reduced to 60\% and 50\%. We defined the boundary layer width
$w$ as the intersection abscissa of the linear fits for the bulk
and near-the-wall parts of the velocity profile. It is clear from
the plot that $w$ is independent of $\Omega$ for each solution and
is seen to depend only on the solution viscosity, whereas we obtained
the fit $w\propto\eta^{0.26\pm0.05}$ (upper inset of Fig.~\ref{f.24}).
Fluctuations of the azimuthal velocity are small near the upper plate,
reach a maximum at $z/d\simeq0.1$, and start to decrease at larger
$z$. Again, such distribution of $V_{\theta}^{rms}$ along $z$ is
reminiscent of the velocity fluctuations in the boundary layer of
turbulent flows of a Newtonian fluid~\cite{landau,Tritt}. An example
of the profile of the rms of the azimuthal velocity $V_{\theta}^{rms}$,
obtained at various $\Omega$ in the setup $2$ with height $d=1$cm,
is shown in Fig.~\ref{f.25}. Well-defined peaks appear at the same
position, which does not depend on $\Omega$, and match the boundary
layer thickness $w$ found with the previous analysis of the average
velocity profiles, and displayed in Fig.~\ref{f.24}.

\begin{figure}
\begin{centering}\centering \includegraphics[width=14cm]{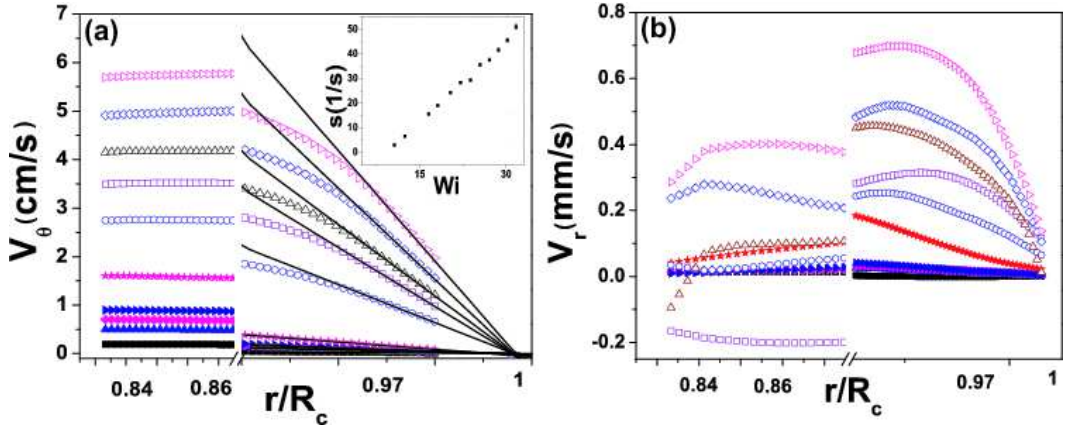} \end{centering}

\caption{Near-wall profiles of the velocity in a horizontal plane at $z=d/2$,
for various values of $Wi$ in setup 2, with $65\%$ saccharose polymer
solution. (a) Radial profiles of the tangential velocity component
$V_{\theta}$. The full lines are linear fit functions. The inset
shows the dependence of the slope of the linear fit functions, $s$,
on $Wi$. (b) Radial profiles of the radial velocity component $V_{r}$.
The symbols are: full squares-$Wi=2.4$, full up triangles-$Wi=4$,
full diamonds-$Wi=5.9$, full right triangles-$Wi=7$, full stars-$Wi=10.6$,
empty circles-$Wi=16.6$, empty squares-$Wi=20.3$, empty up triangles-$Wi=23.8$,
empty diamonds-$Wi=23.8$, empty right triangles-$Wi=30.2$.\label{f.26} }
\end{figure}

We also measured by PIV radial profiles of both the azimuthal and
the radial velocity in a horizontal plane at $z=d/2$ in the vicinity
of the vertical wall of the cell, shown in Fig.~\ref{f.26}(a,b).
As shown in Fig.~\ref{f.26}a, the azimuthal velocity component $V_{\theta}$
can be fitted by a linear function in the vicinity of the wall. This
agrees with the requirements of incompressibility, boundary condition
for the velocity and smoothness of the velocity field, as pointed
out in Ref.~\cite{Turitsyn}.

In the regime of elastic turbulence the slopes of the linear fit functions
increase monotonically with $Wi$ (see the inset in Fig.~\ref{f.26}a).
Using the fits one can determine the width of the vertical (close
to the exterior cylindrical wall) boundary layer, $w_{\theta}$, of
the azimuthal velocity.

\begin{figure}
\begin{centering}\centering \includegraphics[width=9cm]{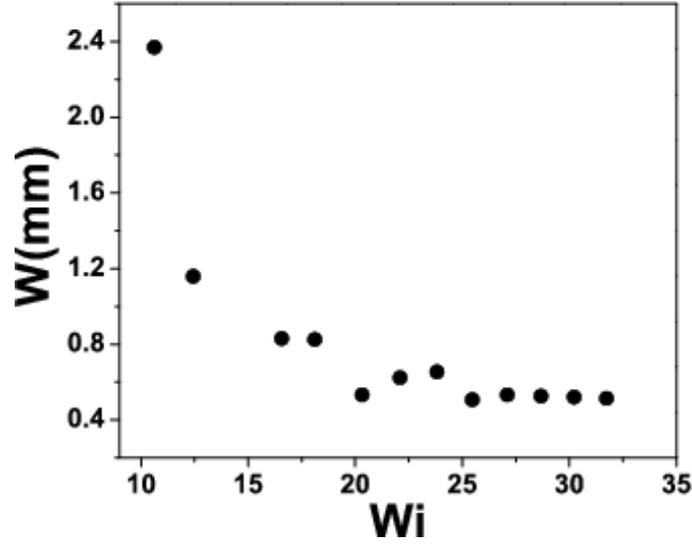} \end{centering}

\caption{Dependence of the width of the vertical boundary layer, $w_{\theta}$
(defined via the width of the linear part of the near wall $PIV$
profiles of the tangential velocity component) on $Wi$. Data were
collected in setup $2$.\label{f.27} }
\end{figure}

As shown in Fig.~\ref{f.27}, the width of the vertical boundary
layer, $w_{\theta}$, decreases abruptly above the onset of elastic
instability and saturates in the regime of elastic turbulence, for
large values of $Wi$. Unfortunately the limited accuracy of measurements
of the radial velocity component near the wall (note that radial velocities
are about two orders of magnitude smaller than the azimuthal ones)
did not allow us to assess beyond doubt the functional behavior of
the radial profiles of radial velocity component near wall, presented
in Fig.~\ref{f.26}b.

\begin{figure}
\includegraphics[width=9cm]{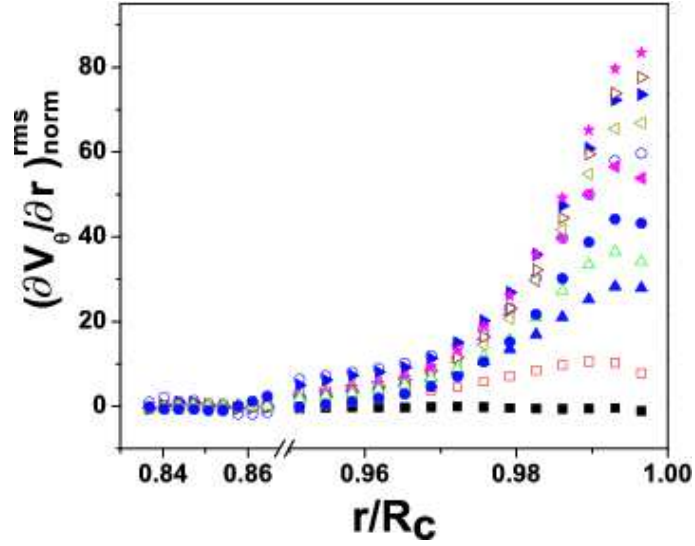}

\caption{Normalized radial profiles of rms of the azimuthal velocity gradients,
$({\partial V_{\theta}/\partial r})_{norm}^{rms}$ in the vicinity
of the vertical wall, for various values of $Wi$ (color online):
full squares-$Wi=10.6$, empty squares-$Wi=12.4$, full up triangles-$Wi=16.5$,
empty up triangles-$Wi=18.1$, full circles-$Wi=20.3$, full left
triangles-$Wi=22$, empty circles-$Wi=23.8$, empty left triangles-$Wi=25.4$,
full right triangles-$Wi=27$, empty right triangles-$Wi=28.6$, stars-$Wi=30.2$.
Data were collected using PIV in setup~$2$ with a $65\%$ saccharose
polymer solution, in a horizontal plane at $z=d/2$.\label{f.28}}
\end{figure}

In order to improve the spatial resolution of the velocity field measurements
near the wall, and particularly to resolve the horizontal boundary
layer by looking at the rms of the radial gradient of the azimuthal
velocity, $(\partial V_{\theta}/\partial r)^{rms}$, the field of
view of the PIV measurements was reduced down to $10mm$ in the radial
direction. The radial profiles of $(\partial V_{\theta}/\partial r)_{norm}^{rms}$
near the wall for various values of $Wi$ in the elastic turbulence
regime are shown in Fig.~\ref{f.28}. As in the case of the average
velocity in the vertical plane (Fig.~\ref{f.24}), the bulk part
of the radial dependence fitted by a line was subtracted for each
curve. Well-defined peaks, positioned independently of $Wi$, can
easily identified inside the velocity boundary layer. We would like
to point out that the value of $(\partial V_{\theta}/\partial r)^{rms}$
at the maximum exceeds up to two orders of magnitude its bulk value
(see Fig.~\ref{f.16}(b)).

\begin{figure}
\includegraphics[width=9cm]{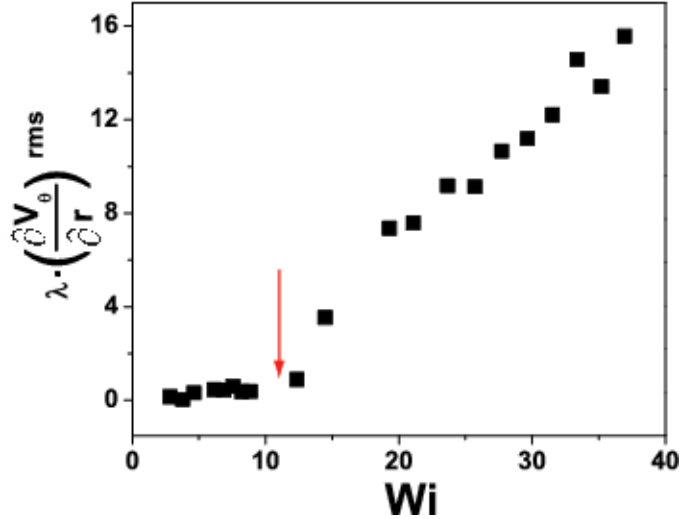}

\caption{Scaled rms of the velocity gradients in the boundary layer vs $Wi$.
The arrow indicates the onset of the primary elastic instability.\label{f.29} }
\end{figure}

The same data were also used to determine the average value of $(\partial V_{\theta}/\partial r)^{rms}$
in the boundary layer. The latter was obtained from widths of distributions
of the slopes of the radial dependence of the azimuthal velocity in
the boundary layer region, for each value of $Wi$. These average
values of $(\partial V_{\theta}/\partial r)^{rms}$ scaled by $\lambda$,
in the vertical boundary layer of the setup $2$ with cell height
$d=1$cm are shown in Fig.~\ref{f.29} as function of $Wi$. A clear
increase of $(\frac{\partial V_{\theta}}{\partial r})^{rms}\cdot\lambda\gg1$
in the boundary layer, in contrast to its saturation value in the
bulk, $(\frac{\partial V_{\theta}}{\partial r})^{rms}\cdot\lambda\approx2$,
was found in the same range of $Wi$ (see further discussion).

\subsection{Discussion: role of the elastic stresses in elastic turbulence in
a finite size cell.\label{sub:Discussion}}

Let us summarize our different experimental observations, discuss
some of their implications, and delineate a comprehensive picture
of the phenomenon called elastic turbulence.

\textbf{(i)} The rms of the vorticity (and that of velocity gradients)
saturates in the bulk of the elastic turbulent flow, leading to the
saturation of the elastic stress. The saturation of $\omega_{rms}$
at high values of $Wi$ in the elastic turbulence regime, observed
in the data in Fig.~\ref{f.16}, gives us the possibility to test
which of the two theoretical mechanisms is responsible for it: either
the feedback reaction of polymer molecules on the flow~\cite{lebedev}
or the nonlinearity of polymer elasticity~\cite{chertkov}. The local
Weissenberg number can be estimated from Fig.~\ref{f.16} as $Wi'=\omega_{rms}\cdot\lambda\simeq2$,
which exceeds the unit value predicted by the model with feedback
of linear molecule elasticity, discussed by~\cite{Volodya,Volodya2}.
We conclude that the nonlinearity of the polymer elasticity may also
contribute to the saturation of the elastic stresses~\cite{chertkov}.
The saturation of the elastic stress in the bulk also naturally explains
the power-law behavior of the average injected power, $\frac{\bar{P}}{\bar{P}_{lam}}\sim Wi^{0.49}$,
seen in the plot in Fig.~\ref{f.3}. Indeed, the injected power is
proportional to the torque, $P=T\cdot\Omega$. The latter is just
the shear stress averaged over the upper plate, and its growth above
the instability threshold occurs solely due to the elastic stress,
$\bm{\tau}_{p}$,~\cite{stretch}. In the elastic turbulence regime,
however, the theory predicts $\tau_{p}\sim\eta/\lambda$. On the other
hand, it was found that, due to shear thinning of the solution used,
$\lambda\sim\Omega^{-0.3}$. Thus, the theoretically expected growth
of the mean injected power with respect to its laminar value, $\bar{P}_{lam}$,
should be solely due to the elastic stress and have the following
power-law scaling: $\frac{\bar{P}}{\bar{P}_{lam}}\sim\tau_{p}\sim Wi^{0.43}$
which is rather close to the scaling we observed experimentally.

Next, we made three main observations \textbf{(ii)}-\textbf{(iv)}
below, which extend further the formal similarity with hydrodynamic
turbulence, despite the distinctively different underlying mechanism
for intermittency. Our findings in the statistics and structure of
the velocity field have to be related to the finite size of the experimental
vessel, whereas the current theory describes elastic turbulence in
an unbounded domain~\cite{Volodya,Volodya2}. One can expect to find
new effects in statistical and scaling behavior, similar to those
recently studied both theoretically and experimentally for a passive
scalar advected by a smooth random flows in bounded geometries~\cite{chertkov2},
such as that in a channel considered by~\cite{Teo}.

As was already pointed out in the Introduction, there is a deep analogy
in the statistical behavior of elastic turbulence, of the fast magneto-dynamo,
and for the passive scalar dispersion in the Batchelor regime. As
was shown in Section~\ref{sub:ElasticTheory}, the elastically driven
random flow can be considered as turbulence of the elastic stress
field; all its statistical properties can be explained by the interplay
of (a) stretching and folding of fluid elements carrying elastic stresses
on Lagrangian trajectories in the Batchelor regime, (b) dynamo effect
leading to the saturation of stresses in the bulk flow, and (c) homogeneous
linear damping of the stress. Thus, since the dynamics of the elastic
stress tensor can be reduced to the dynamics of the vector field $\mathbf{B}$,
this scenario is very close to that found in the fast magneto-dynamo
except that (c) is replaced by the diffusive dissipation of the magnetic
field. On the other hand, the passive scalar problem differs from
elastic turbulence (besides the fact that in this case a vector is
replaced by a scalar) in the item (b), since in the former the feedback
reaction on the flow is absent. However, in spite of the fact that
there is no exact similarity between these three problems, the main
ingredient (a) gives a basis for the similarity in statistical and
scaling behavior of these systems. Taking into account the similarity,
we can extend it to statistical properties of elastic turbulence in
a finite size vessel.

\textbf{(ii)} The rms of the velocity gradients $\left(\frac{\partial V_{\theta}}{\partial r}\right)^{rms}$
or $\omega^{rms}$, and thus the elastic stress, grows linearly with
$Wi$ in the boundary layer (see Fig.~\ref{f.29}), near the driving
disk. The rms of the velocity gradients saturates in the bulk (see
Fig.~\ref{f.16}). Moreover, the values in the boundary layer are
one to two orders of magnitude larger than in the bulk. All this implies
that a boundary layer is formed, with an excess of the elastic stress
is concentrated in it.

\textbf{(iii)} The PDFs of the injected power $\delta P_{\Omega}/P_{\Omega}^{rms}$,
at either constant angular speed or torque, show skewness and exponential
tails, which both indicate intermittent statistical behavior. Also
the PDFs of the normalized accelerations, which can be related to
the statistics of velocity gradients via the Taylor hypothesis, exhibit
well pronounced exponential tails. We can match this facts and the
emergence of the boundary layer with the following scenario: the forcing
from the moving boundary (the upper disk) produces an accumulation
of the elastic stress in the boundary layer due to a constant flux
of momentum from the upper boundary. Elastic stress is intermittently
injected into the bulk by random release events. Whenever that happens,
the power on the disk decreases. Constant build-up and sudden release
would thus explain the skewness of the PDF toward small values of
the injected power, while intermittency would be the origin of the
observed exponential tails. The exponential tails in the PDF of the
velocity gradients and accelerations (Fig.~\ref{velstat}-\ref{f.19})
can be also explained in the same way, except that for them the symmetry
of the PDF is preserved due to injections from both the upper and
lower boundary layers.

However, the most surprising and striking feature of elastic turbulence
observed is \textbf{(iv)}, the presence of a velocity boundary layer
and the emergence of a new length scale. Since elastic turbulence
is a turbulence of an elastic stress field, one can expect primarily
the appearance of a boundary layer in the elastic stress field. We
observe in addition the presence of a velocity boundary layer, which
just reflects some of the features of the boundary layer of the elastic
stresses (rms of the velocity gradients). The measured thickness of
the velocity boundary layer is much smaller than the vessel size.
To argument further on it, we note that the picture presented in point
\textbf{(iii)} is in close analogy with the turbulent advection of
a passive scalar in the Batchelor regime in a finite channel flow,
where the excess of tracer from the boundary layer is intermittently
injected into the bulk~\cite{chertkov2,Teo}. There the boundary
layer for a passive scalar (mixing boundary layer) results from zero
boundary conditions for the flow velocity components along and towards
the wall. This leads to less effective advection of the passive scalar
near the wall, causing an accumulation in the near-wall region, and
thus a boundary layer in which the statistics of the scalar is different
than that in the bulk flow. In the case of the passive scalar, the
mixing boundary layer and the new length scale appear due to the existence
of a small parameter, the inverse Peclet number, $Pe^{-1}=(D/V^{rms}L)\ll1$,
where $D$ is the diffusion constant, and $L$ is the vessel size.
Then, as predicted theoretically~\cite{chertkov2} and confirmed
experimentally~\cite{Teo}, the mixing boundary layer width is found
to scale as $w\propto Pe^{-1/4}\propto D^{1/4}$. Intermittent ejection
of the tracer into the bulk, analogous to the intermittent injection
of elastic stresses in elastic turbulence, reduces the effectiveness
of mixing, increases the mixing length, and causes stronger (power-law)
dependence of the mixing length on $Pe$ than in an unbounded vessel
(in which dependence should be logarithmic).

Carrying on the analogy with the passive scalar problem, one can suggest
that also here a small parameter of the problem defines a new length
scale. Since diffusion does not play any significant role in elastic
turbulence, we suggest that

\begin{equation}
N\equiv\left(\frac{\partial V_{\theta}}{\partial r}\right)_{bulk}^{rms}\left/\left(\frac{\partial V_{\theta}}{\partial r}\right)_{b.l.}^{rms}\right.\ll1\label{eq:NboundaryLayer}\end{equation}
 can be considered as the relevant small parameter. This non-uniform
distribution of the elastic stresses should result in, and be directly
related to, a non-uniform distribution of polymer stretching: the
polymer molecules should be stretched considerably more in the boundary
layer than in the bulk. We have verified this prediction in a separate
experiment on a single molecule in a random flow~\cite{FlyingSaucer}.
Thus, the width of the boundary layer for the elastic stress, based
on these considerations and on the experimental observations, scales
like $w/L\propto N^{\mu}\eta^{1/4}$, with $\mu$ an exponent to be
determined. While it is appealing to note that the exponent of $\eta$
(Fig.~\ref{f.24}) is the same as that of $1/Pe$ in the previous
case, we are not really able to justify it theoretically. On one hand,
in contrast with the passive scalar problem, where dissipation due
to diffusion introduces the natural length scale, elastic stresses
in elastic turbulence are dissipated via linear damping with the polymer
relaxation time, $\lambda$. Since $\lambda\propto\eta$, $\eta$
has good reason to appear in the scaling relation for $w$, but we
cannot guess a value of its exponent. On the other hand, the suggested
scaling has a serious flaw: while we have seen that $\left(\frac{\partial V_{\theta}}{\partial r}\right)_{bulk}^{rms}$
saturates in the regime of well developed elastic turbulence (Fig.~\ref{f.16}b),
the suggested scaling law bears a dependence on $\Omega$ via $\left(\frac{\partial V_{\theta}}{\partial r}\right)_{b.l.}^{rms}$
(see Fig.~\ref{f.29}), whereas we found experimentally that $w$
is independent of it. These two issues need to be addressed by a forthcoming
theory.

Finally, \textbf{(v)}, the scaling of the structure functions of the
vorticity, velocity gradients, and injected power is found to be the
same of that of a passive scalar advected by an elastic turbulent
velocity field (see subsection~\ref{sub:Flow-structure}), fact which
further reinforces the analogy in the statistics and scaling of the
passive scalar and of the elastic stresses due to advection by a smooth
random flow.

\section{Conclusions\label{sec:Conclusions}}

The main message of this study is a surprising similarity in scaling,
statistics, and structure of both the elastic stresses and the passive
scalar in elastic turbulence in a finite size vessel, in spite of
the important difference in the dynamo effect.

First of all, we have confirmed that the flow is smooth and essentially
dominated by large scale structures: the velocity power spectra decays
as $P\sim k^{-3.3}$, which suggests that the power of fluctuations
of velocity gradients scales as $k^{-1.3}$ (see Sec.~\ref{sub:Flow-structure}).
As explained in Sec.~\ref{sub:ElasticTheory}, the main contribution
to the fluctuations of the velocity gradients and of the velocity
differences at all scales comes from the largest eddies, having dimensions
of the whole system (the gap between plates). This conclusion has
an immediate implication for the mixing in the flow: it should result
in the same type of patterns and in the same statistics observed in
a randomly varying in time flow with a linear velocity field (locally
uniform rate of strain), $\mathbf{V}(\mathbf{r},t)=\mathbf{V_{0}}(t)+\frac{\partial\mathbf{V}(t)}{\partial\mathbf{r}}\cdot\left(\mathbf{r}-\mathbf{r}(0)\right)$,
which approximates well the velocity field in elastic turbulence~\cite{taylorhypoth}
(\textbf{$r$} is the position vector).

Looking at more detail into the local properties of the velocity field,
we have evidenced a saturation of the rms of the vorticity (or velocity
gradients), which should lead to the saturation of the elastic stresses
in the bulk. This agrees with theoretical predictions, save the fact
that the saturation level found is about twice than what theoretically
predicted~\cite{Volodya,Volodya2}. The discrepancy can be attributed
to nonlinear polymer elasticity~\cite{chertkov}. The observed saturation
is however remarkably consistent with the scaling of the averaged
injected power with $Wi$.

The presence of a boundary layer for stress and velocity was also
evidenced and characterized: the rms of the velocity gradients (and
therefore the elastic stresses) grow linearly with $Wi$ within it,
and exceed the corresponding value in the bulk by one-two orders of
magnitude, suggesting that the elastic stresses are accumulated near
the wall and are intermittently injected into the bulk. The view is
confirmed by the properties of the PDFs of the injected power and
by the exponential tails of PDFs of the normalized accelerations.
This leads to a model for the dynamics of elastic turbulence, in which
elastic stress is introduced into the fluid by the driving boundary,
accumulates in the boundary layer and is intermittently injected into
the bulk of the flow. The situation is entirely similar to the trapping
of a passive scalar in the mixing boundary layer in a finite size
vessel, and to its intermittent injection into the bulk. The non-uniform
distribution of the elastic stresses, which causes non-uniform polymer
stretching is currently verified experimentally~\cite{FlyingSaucer}.

Universality and scaling of the boundary layer thickness are also
investigated. The emergence of a relevant scale for it, based on properties
measured from the velocity profile, is the most unexpected and striking
result, since according to the theory of elastic turbulence in unbounded
domains, there would be no other length scale besides the dissipation
length or the vessel size. The measured boundary layer is much thinner
than the vessel size, and the scaling of its width $w$ with $\eta$
was found experimentally. A small parameter, which is related to the
polymer extension ratio in the bulk and in the boundary layer, was
identified in this problem.

Finally, the similarity between elastic turbulence and Batchelor regime
advection of passive scalar is reinforced by the scalings of structure
functions of the vorticity, velocity gradients, and injected power,
which were found to be the same as that for a passive scalar advected
in a velocity field produced by elastic turbulence.

The experimental results presented call for further development in
the theory of elastic turbulence in a container of finite size. Particular
attention should be devoted to what causes the scaling of the boundary
layer width, the scaling of structure functions and rms of the injected
power, and the saturation level of the elastic stresses in the bulk
flow.

\section*{Acknowledgments}

We thank M. Chertkov and V. Lebedev for numerous discussions and suggestions.
This work was supported by the grants of Israel Science Foundation,
Binational US-Israel Foundation, and by the Minerva Center for Nonlinear
Physics of Complex Systems.

\end{document}